\documentclass[intlimits,twoside,a4paper]{article}
\usepackage{wrapfig}
\usepackage{graphicx}
\usepackage{amsmath,amssymb}
\usepackage{color}
\usepackage[T2A]{fontenc}
\usepackage[cp1251]{inputenc}
\DeclareMathOperator{\Sp}{Sp}

\usepackage[eqsecnum]{cmpj3}

\newcommand{\be}{\begin{equation}}
\newcommand{\ee}{\end{equation}}
\newcommand{\bea}{\begin{eqnarray}}
\newcommand{\eea}{\end{eqnarray}}

\issue{2018}{21}{3}{33702}
\doinumber{10.5488/CMP.21.33702}

\title[Deformation effects]{Deformation effects in glycinium phosphite ferroelectric}
\author[I.R. Zachek, R.R. Levitskii, A.S. Vdovych]{I.R. Zachek\refaddr{label1}, R.R. Levitskii\refaddr{label2}, A.S. Vdovych\refaddr{label2}}
\addresses{\addr{label1} Lviv Polytechnic National
University, 12 Bandera St., 79013 Lviv, Ukraine
\addr{label2} Institute for Condensed Matter Physics of the
National Academy of Sciences of Ukraine,\\ 1 Svientsitskii St.,
79011 Lviv, Ukraine }

\date{Received June 13, 2018}

\begin{document}

\maketitle

\begin{abstract}

To study the effects appearing under mechanical stresses, we have used the model  of  glycinium phosphite ferroelectric, modified by taking into account the piezoelectric coupling of ordering structure elements with lattice strains. In the frames of two-particle cluster approximation,
the components of polarization vector and static dielectric permittivity tensor of the crystal, as well as its piezoelectric and thermal characteristics  are calculated. 
Influence of shear stresses, hydrostatic and uniaxial pressures on the phase transition and physical characteristics of the crystal is studied.

\keywords ferroelectrics, phase transition, dielectric permittivity, piezoelectric characteristics, mechanical stress
\pacs 77.22.-d, 77.22.Ch, 77.22.Ej, 77.65.-j, 77.80.Bh

\end{abstract}

\section{Introduction}

Investigation of properties of ferroelectric materials under external pressures is one of the actual problems in condensed matter physics.
Considerable part of these materials are compounds, where the order-disorder phase transitions take place. Their behaviour is properly described by quantum-statistical models. The most known examples of this class of materials are crystals with hydrogen bonds.

The application of external pressures with different symmetry is an effective means for a continuous change of geometric characteristics of hydrogen bonds, which let us more profoundly study the role of these bonds in mechanisms of phase transitions and in dielectric response of these crystals. Piezoelectrics in paraelectric phase form a large part of the ferroelectric compounds with hydrogen bonds.
Application of shear stresses make it possible to study the role  of piezoelectric interactions in the phase transitions and in forming the piezoelectric, elastic and dielectric characteristics of these crystals.

It is worth noting that the application of pressures and shear stresses  is a very important tool in investigating the  ferroelectric crystals with a complex structure of effective dipole moments. Such compounds often comprise several sublattices of effective dipoles, which  are not always mutually parallel. 

The effect of hydrostatic pressure on the KH$_2$PO$_4$ family crystal is the most fully studied experimentally among the compounds with hydrogen bonds. History of these investigations exceeds forty years. During this period, a great amount of experimental data \cite{Samara925,Tibbals849,Nelmes1041,Peercy2725,Schmidt839,Nelmes125,Nelmes87} was accumulated. It was determined that this pressure significantly influences  the phase transitions in these crystals while their physical characteristics  noticeably change. 
Unfortunately, the effect of pressures with another symmetry on the above mentioned crystals is much less studied. It is only known that the phase transition temperature in  KH$_2$PO$_4$ and KD$_2$PO$_4$ noticeably decreases \cite{Stadnyk317} under uniaxial pressure  $p_{zz}=-\sigma_{zz}$ along the axis of spontaneous polarization.

Theoretical description of the behaviour of ferroelectric compounds with hydrogen bonds, including KH$_2$PO$_4$ type, was based on the proton ordering model \cite{Blinc204,deGennes132,Blinc430}. For the first time, a modification of this model on the case of deformed crystals was accomplished in \cite{Blinc701,Torstveit4431}, where within four-particle cluster approximation there was originally achieved an agreement of theory with experimental data for the pressure dependences of the phase transition temperature, Curie-Weiss constant and saturation polarization.
It should be noted that deformational effects in these theories were taken into account semiphenomenologically, assuming that parameters of tunneling, short-range and effective long-range  interactions are functions of the distance between equilibrium positions of proton on the hydrogen bond $\delta$ linearly decreasing with pressure. The value of effective dipole moment of the crystal was also considered to be proportional to the distance $\delta$.

Later on in the papers by Stasyuk and Biletskii \cite{Stasyuk705,StasyukITF8393R} there was originally realized a microscopic substantiation of the methods that take account of the lattice strains with different symmetry in the proton ordering model. To obtain an effective model of a deformed crystal of KH$_2$PO$_4$ type, the microscopic Hamiltonian was used that takes into account its proton and lattice subsystems and also piezoelectric and electrostrictive interactions between pseudospin variables, acoustic and optical phonons as well as takes account of cubic anharmonism.
It was shown by means of separation of a lattice strain in the mean field approximation that the effect of this strain reduces to the appearance of 
 internal fields that depend on the lattice symmetry. These fields can include contributions connected both with the internal piezoeffect and with the pseudospin interaction renormalized by electrostriction. 
As a result, according to the model, proposed by the authors, the application of external pressure to the crystal  leads to the appearance of an additional internal field, which is linear on the strains and the mean values of pseudospins describe the positions of protons on the hydrogen bonds. The energies of these configurations are also considered to be linearly dependent on the strains. The effect of pressure of different symmetry on these energies is studied. 
In  \cite{Stasyuk705,Stasyuk567,Stasyuk553}, the model of a deformed crystal was used for a description of the effects caused by the action of symmetrized stress  $\sigma_{xx}-\sigma_{yy}$ on the KH$_2$PO$_4$ type ferroelectrics. As a result of calculations, there was predicted a possibility of the phase transition into the new hypothetic phase with monoclinic symmetry induced by such a stress.

Within the same approach in \cite{Duda129,Moina8530,Moina731,Romanyuk502,kn2009}, there was realized a consistent description of the effects of external hydrostatic and uniaxial  $p_{zz}=-\sigma_{zz}$ pressures on the physical characteristics of several KH$_2$PO$_4$ type ferroelectric crystals. On the basis of the same model of a deformed crystal in \cite{Stasyuk6198,Stasyuk213}, there was developed a microscopic theory of the effects of stress $\sigma_{xy}$ and electric field $E_z$ on the phase transition and physical characteristics of KH$_2$PO$_4$ family compounds that takes account of piezoelectric coupling. 

\looseness=-1 This year eighty years passes since the discovery of ferroelectricity of KH$_2$PO$_4$ crystal. During this period, a large amount of papers, reviews and monographs (see \cite{kn2009}) was devoted to the investigation of the phase transition  and physical characteristics of KH$_2$PO$_4$ family compounds.  The most noticeable peculiarity of the physics of these materials is a close cooperation between theory and experiment, which is an important reason for the progress achieved at that time in the microscopic description of their properties. 
It is worth noting that exactly the model of deformed ferroelectric crystals proposed in \cite{Stasyuk705,StasyukITF8393R} ensured a considerable progress in the future development of their microscopic theories. Following the publication of papers  \cite{Duda129,Moina8530,Moina731,Romanyuk502,kn2009,Stasyuk6198,Stasyuk213} in \cite{Levitskii4703}, the piezoelectric coupling was taken into account during the study of the effect of hydrostatic pressure on the physical characteristics of KH$_2$PO$_4$ type crystals. Later on, within the models of deformed ferroelectric crystals, the investigations  of the effect of hydrostatic pressure  on the phase transition  and on the physical characteristics  of quasionedimensional CsH$_2$PO$_4$ type ferroelectrics \cite{lev3}, monoclinic RbD$_2$PO$_4$ \cite{zac}, as well as  RbHSO$_4$  \cite{zac2} were also carried out.

Late in the twentieth century, the ferroelectricity and a unique sensitivity of the crystal glycinium hydrogenphosphite NH$_3$CH$_2$COOH$\cdot$H$_2$PO$_3$ (GPI) to a transverse electric field  $E_z$ was discovered.
GPI is a very interesting compound due to  the combination of structural elements typical of different classes of crystals. Very important, it contains the covalently bonded phosphite HPO$_3$ groups linked through the hydrogen O-H$\ldots$O bonds, thus forming the chains running along the $c$-axis.  Such structural components are usual for non-organic ferroelectric materials, in particular, for crystals of KH$_2$PO$_4$ family.  Besides, there are four organic glycinium groups NH$_3$CH$_2$COOH in GPI unit cell that are linked by four additional hydrogen bonds with phosphite HPO$_3$ groups belonging to two different phosphite chains. 
At room temperature, GPI crystalizes in a monoclinic P2$_1/a$ space group, which transforms to P2$_1$ symmetry \cite{Averbuch1993,Shikanai2002,Taniguchi2003} below the structural phase transition temperature. As was mentioned above, these crystals belong to  the ferroelectric crystals with hydrogen bonds \cite{dac,Baran1996}.  There are two structurally inequivalent types of hydrogen bonds of different length,  $\approx2.48$~{\AA} and  $\approx2.51$~{\AA}. Proton ordering on the hydrogen bonds \cite{Shikanai2002,Taniguchi2003} causes antiparallel orientation of the components of dipole moments along the crystallographic axes \textit{a} and \textit{c} in the neighbouring chains. The strains of tetrahedra  HPO$_3$ and the corresponding  components of dipole moments along the \textit{b}-axis in the chains cause the total dipole moment along the \textit{b}-axis. As a result, at temperature 225~K, the GPI crystal passes to the ferroelectric state with spontaneous polarization perpendicular to the chains of hydrogen bonds.

The experiment, carried out in \cite{Stasyuk2003}, revealed anomalies of permittivity  $\varepsilon_{zz}$ in the phase transition region at $E_z\neq0$ and a decrease of the phase transition temperature proportional to $E_z^2$. For the first time, the explanation of the revealed effects was given on the basis of phenomenologic theory \cite{Stasyuk2004} and within microscopic approach \cite{Stasyuk2003,Stasyuk2004Ferro}. Unfortunately, a complete quantitative description of these experimental data was not obtained. Later on, this problem was solved in  \cite{Zachek_CMP2017} on the basis of the model of deformed  GPI crystal proposed in  \cite{Zachek_PB2017}, which is generalization of the proton  model by Stasyuk and Velychko \cite{Stasyuk2003,Stasyuk2004Ferro}.

The model of deformed  GPI crystal \cite{Zachek_PB2017} made it possible to correctly describe polarization and components of dielectric permittivity tensor for a mechanically free and clamped crystal, its piezoelectric, elastic characteristics and heat capacity, influence of longitudinal field  $E_y$ \cite{Vdovych_UJP350},  hydrostatic pressure \cite{Zachek_CMP2017_p} and uniaxial pressures \cite{Zachek_JPS2017} on these characteristics, as well as relaxation phenomena \cite{Zachek_CMP2018dyn}.

In the present paper, the GPI model \cite{Zachek_PB2017,Zachek_CMP2017} is modified for the case of a decreasing  symmetry under shear stresses   $\sigma_{yz}$ and  $\sigma_{xy}$.  The effects of different mechanical stresses  on the phase transition, dielectric and piezoelectric  characteristics of this crystal are studied.

\section{Hamiltonian of the model}

\begin{figure}[!b]
	\begin{center}
		\includegraphics[scale=0.65]{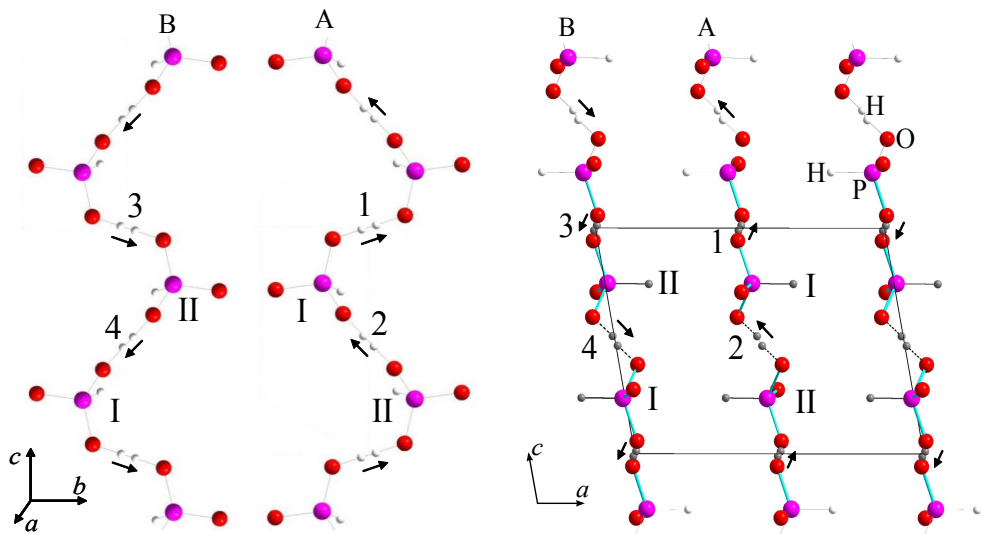}
	\end{center}
	\caption{(Colour online) Orientations of vectors ${\bf d}_{qf}$ in the primitive cell
		in the ferroelectric phase \cite{Zachek_PB2017}.} \label{struktura}
\end{figure}

We consider the system of protons in GPI, localized on O-H$\ldots$O bonds between phosphite groups HPO$_{3}$, which form zigzag chains along the crystallographic $c$-axis of the  crystal \cite{Zachek_PB2017,Zachek_CMP2017} (figure~\ref{struktura}). For a better understanding of the model, only phosphite groups are shown in the figure.
Dipole moments ${ \bf d}_{qf}$ $(f=1,\dots,4)$ are ascribed to the protons on the bonds.  In the ferroelectric phase, the dipole moments compensate each other  (${\bf d}_{q1}$ with ${\bf d}_{q3}$, ${\bf d}_{q2}$ with ${\bf d}_{q4}$) in directions  $Z$ and $X$ ($X\perp (b,c)$, $Y \parallel b$, $Z \parallel c$), and simultaneously supplement each other in direction $Y$, creating spontaneous polarization. 
Pseudospin variables
$\frac{\sigma_{q1}}{2},\dots,\frac{\sigma_{q4}}{2}$ describe  reorientation of the dipole moments of the base units: ${\bf d}_{qf} = {\boldsymbol{\muup}}_f \frac{\sigma_{qf}}{2}$.
Mean values  $\langle \frac{\sigma}{2}\rangle = \frac12
(n_a-n_b)$ are connected with differences in the occupancy of the two possible molecular positions, $n_a$ and $n_b$.

Herein below, for components of vectors and tensors for convenience we use the notations $1$, $2$ and $3$ instead of $x$, $y$ and $z$.
Hamiltonian of proton subsystem of GPI, which takes into account the short-range and long-range interactions and the applied electric fields $E_1$, $E_2$, $E_3$ along positive directions of the Cartesian axes    $X$, $Y$ and $Z$ can be written in such a way:
\bea
&& \hat H= N U_{\text{seed}} + \hat H_{\text{short}} + \hat H_{\text{long}}+ \hat H_{E}\,,  \label{H}
\eea
where $N$  is the total number of primitive cells.
The term $U_{\text{seed}}$ in (\ref{H})  is the ``seed'' energy, which relates  to the heavy ion sublattice and does not depend explicitly on the configuration of the proton subsystem. It includes the elastic, piezoelectric and dielectric parts, expressed in terms of electric fields
$E_i$ $(i=1, 2, 3)$  and strains  $\varepsilon_j$ $(j=1,\ldots,6)$. 
\bea
&& U_{\text{seed}}= v\left[\frac{1}{2}\sum\limits_{j,j'=1}^6c_{jj'}^{E0}(T)\varepsilon_j \varepsilon_{j'} - \sum\limits_{i=1}^3 \sum\limits_{j=1}^6 e_{ij}^0 \varepsilon_j E_i  -  \sum\limits_{i,i'=1}^3 \frac{1}{2}  \chi_{ii'}^{\varepsilon 0}E_iE_{i'}\right].
\eea
Parameters $c_{jj'}^{E0}(T)$, $e_{ij}^0$, $\chi_{ii'}^{\varepsilon 0}$ are the so-called ``seed'' elastic constants, coefficients of
piezoelectric stresses and dielectric susceptibilities, respectively, $v$ is the volume of a primitive cell.
Matrices $c_{jj'}^{E0}$, $e_{ij}^0$, $\chi_{ii'}^{\varepsilon 0}$ are given by:
\be
\hat{c}_{jj'}^{E0} = \left(
\begin{array}{cccccc}
	c_{11}^{E0}	& c_{12}^{E0} & c_{13}^{E0} & 0 & c_{15}^{E0} & 0 \vspace{0.8mm}\\
	c_{12}^{E0}	& c_{22}^{E0} & c_{23}^{E0} & 0 & c_{25}^{E0} & 0 \vspace{0.8mm}\\
	c_{13}^{E0}	& c_{23}^{E0} & c_{12}^{E0} & 0 & c_{35}^{E0} & 0 \vspace{0.8mm}\\
	0	& 0 & 0 & c_{44}^{E0} & 0 & c_{46}^{E0} \vspace{0.8mm}\\
	c_{15}^{E0}	& c_{25}^{E0} & c_{35}^{E0} & 0 & c_{55}^{E0} & 0 \vspace{0.8mm}\\
	0	& 0 & 0 & c_{46}^{E0} & 0 & c_{66}^{E0}
\end{array} \right),
\begin{array}{c}
	\hat{e}_{ij}^0 = \left(
	\begin{array}{cccccc}
		0	& 0 & 0 & e_{14}^0 & 0 & e_{16}^0 \\
		e_{21}^0 & e_{22}^0 & e_{23}^0 & 0 & e_{25}^0 & 0 \\
		0	& 0 & 0 & e_{34}^0 & 0 & e_{36}^0
	\end{array} \right), \vspace{1mm}\\
	\hat{\chi}_{ii'}^{\varepsilon 0}  = \left(
	\begin{array}{ccc}
		\chi_{11}^{\varepsilon 0}	& 0 & \chi_{13}^{\varepsilon 0} \\
		0	& \chi_{22}^{\varepsilon 0} & 0 \\
		\chi_{13}^{\varepsilon 0}	& 0 & \chi_{33}^{\varepsilon 0}
	\end{array}  \right).
\end{array}
\ee
In the paraelectric phase, all coefficients $e_{ij}^0 \equiv 0$.

Other terms in  (\ref{H}) describe the pseudospin part of the Hamiltonian.
In particular, the second term in  (\ref{H}) is the Hamiltonian of short-range interactions:
\be
\hat H_{\text{short}} = - 2 \sum\limits_{qq'} \left ( w_{1}\frac{\sigma_{q1}}{2} \frac{\sigma_{q2}}{2} + w_{2}\frac{\sigma_{q3}}{2}\frac{\sigma_{q4}}{2} \right)
\bigl( \delta_{{\bf R}_q{\bf R}_{q'}} + \delta_{{\bf R}_q + {\bf R}_{c},{\bf R}_{q'}} \bigr). \label{Hshort}
\ee
In (\ref{Hshort}),  $\sigma_{qf}$ is the $z$-component of pseudospin operator that describes the state of the $f$-th bond ($f = 1, 2, 3, 4$), in the  $q$-th cell.
The first Kronecker delta corresponds to the interaction between protons in the chains near the tetrahedra HPO$_{3}$ of type ``I'' (figure~\ref{struktura}), where the second  one near the tetrahedra HPO$_{3}$ of type ``II'', ${\bf R}_{c}$ is the lattice vector along $OZ$-axis. Contributions into the energy of interactions between protons near tetrahedra of different type, as well as the mean values of the pseudospins  $\eta_{f}=\langle\sigma_{qf}\rangle$, which are related to tetrahedra of different type, are equal.

Parameters $w_{1}$, $w_{2}$,  which describe the short-range interactions within chains, are expanded linearly into series over strains $\varepsilon_j$:
\bea
&&w_{1} = w^{0} + \sum\limits_{l} \delta_{l}\varepsilon_l +  \delta_{4}\varepsilon_4+  \delta_{6}\varepsilon_6\,, \quad (l=1,2,3,5), \nonumber\\
&&w_{2} = w^{0} + \sum\limits_{l} \delta_{l}\varepsilon_l -  \delta_{4}\varepsilon_4-  \delta_{6}\varepsilon_6. \label{w}
\eea

The third term in  (\ref{H}) describes  the long-range dipole-dipole interactions and indirect  (through the lattice vibrations)  interactions between protons, which are taken into account in the mean field approximation:
\bea
&&\hat H_{\text{long}} = \frac12 \sum\limits_{qq'ff'}  J_{ff'}(qq') \frac{\langle \sigma_{qf}\rangle}{2}\frac{\langle \sigma_{q'f'}\rangle}{2}
- \sum\limits_{qq'ff'}  J_{ff'}(qq') \frac{\langle \sigma_{q'f'}\rangle}{2}\frac{\sigma_{qf}}{2}.\label{Hlong}
\eea
Fourier transforms of interaction constants $J_{ff'} = \sum\nolimits_{q'} J_{ff'}(qq')$ at ${\bf k}=0$ are linearly expanded over the strains $\varepsilon_j$:
\bea
&&J_{\frac{11}{33}} = J^0_{11} + \sum\limits_{l}\psi_{11l}\varepsilon_l \pm \psi_{114}\varepsilon_4\pm \psi_{116}\varepsilon_6\,, \qquad
J_{13} = J^0_{13} + \sum\limits_{l}\psi_{13l}\varepsilon_l + \psi_{134}\varepsilon_4+ \psi_{136}\varepsilon_6\,,\nonumber\\
&&J_{\frac{12}{34}} = J^0_{12} + \sum\limits_{l}\psi_{12l}\varepsilon_l \pm \psi_{124}\varepsilon_4\pm \psi_{126}\varepsilon_6\,, \qquad J_{\frac{14}{23}} = J^0_{14} + \sum\limits_{l}\psi_{14l}\varepsilon_l \pm \psi_{144}\varepsilon_4\pm \psi_{146}\varepsilon_6\,,\nonumber\\
&&J_{\frac{22}{44}} = J^0_{22} + \sum\limits_{l}\psi_{22l}\varepsilon_l \pm\psi_{224}\varepsilon_4\pm \psi_{226}\varepsilon_6\,, \qquad
J_{24} = J^0_{24} + \sum\limits_{l}\psi_{24l}\varepsilon_l + \psi_{244}\varepsilon_4+ \psi_{246}\varepsilon_6.\nonumber
\eea

As a result,  (\ref{Hlong}) can be written as:
\be \hat H_{\text{long}} = N H^{0}  - \sum\limits_q \sum\limits_{f=1}^4 {\cal H}_f \frac{\sigma_{qf}}{2}\,, \label{Hlongs}
\ee
where
\bea &&\ H^{0}  = \sum\limits_{f,f'=1}^{4} \frac18 J_{ff'}\eta_f\eta_{f'}\,, \qquad {\cal H}_f = \sum\limits_{f'=1}^{4}\frac{1}{2}J_{ff'}\eta_{f'}.\label{H0}
\eea

The fourth term in  (\ref{H}) describes the interactions of pseudospins with an external electric field:
\bea
&&\hat H_{E} = -\sum\limits_{qf} {\boldsymbol{\muup}}_{f} {\bf E} \frac{\sigma_{qf}}{2}.\label{H_E}
\eea
Here, ${\boldsymbol{\muup}}_{1}=(\mu_{13}^{x},\mu_{13}^{y},\mu_{13}^{z})$,  ${\boldsymbol{\muup}}_{3}=(-\mu_{13}^{x},\mu_{13}^{y},-\mu_{13}^{z})$,  ${\boldsymbol{\muup}}_{2}=(-\mu_{24}^{x},-\mu_{24}^{y},\mu_{24}^{z})$,  ${\boldsymbol{\muup}}_{4}=(\mu_{24}^{x},-\mu_{24}^{y},-\mu_{24}^{z})$   are the effective dipole moments per one pseudospin.

The two-particle cluster approximation for short-range interactions is used for calculation of thermodynamic characteristics of GPI. In this approximation, thermodynamic potential under stresses $\sigma_j$ is given by:
\bea
&& \hspace{-4ex}  G = N U_{\text{seed}} + NH^0     - N v  \sum\limits_{j=1}^6 \sigma_j \varepsilon_j -  k_{\text B} T  \sum\limits_q \bigg[ 2\ln \Sp \re^{-\beta \hat H^{(2)}_{q}} - \sum\limits_{f=1}^4\ln \Sp \re^{-\beta \hat H^{(1)}_{qf}} \bigg], \label{G}
\eea
where $\beta=\frac{1}{k_{\text B}T}$, $k_{\text B}$ is Boltzmann constant,   $\hat H^{(2)}_{q}$ and $\hat H^{(1)}_{qf}$ are two-particle and one-particle Hamiltonians:
\bea
&& \hat H^{(2)}_{q} = - 2 \left(w_{1} \frac{\sigma_{q1}}{2} \frac{\sigma_{q2}}{2} + w_{2}\frac{\sigma_{q3}}{2}\frac{\sigma_{q4}}{2}\right)
- \sum\limits_{f=1}^4 \frac{y_f}{\beta}  \frac{\sigma_{qf}}{2}\,, \label{H2}\\
&&
\hat H^{(1)}_{qf} = - \frac{\bar y_f}{\beta}\frac{\sigma_{qf}}{2}\,, \label{H1}
\eea
where such notations are used:
\bea
&& \hspace{-4ex} y_f = \beta (  \Delta_1 + {\cal H}_f +  {\boldsymbol{\muup}}_f {\bf E}),  \qquad \bar y_f =  \beta \Delta_f + y_f. \label{yf}
\eea
The symbols $\Delta_f$ are the effective fields created by the neighboring bonds
from outside the cluster. In the cluster approximation, the fields $\Delta_f$ can be determined from the condition of minimum of thermodynamic potential $\partial G/\partial \Delta_f =0$, which gives   the self-consistency
condition, stating that the mean values of the pseudospins $\langle \sigma_{qf} \rangle =\eta_{f}$ calculated with the two-particle
and one-particle Gibbs distribution,  respectively, should coincide:
\bea
\eta_{f} = \frac{\Sp  \sigma_{qf} \re^{-\beta \hat H^{(2)}_{q}}}{\Sp  \re^{-\beta \hat H^{(2)}_{q}}} =
\frac{\Sp  \sigma_{qf} \re^{-\beta \hat H^{(1)}_{qf}}}{\Sp  \re^{-\beta \hat H^{(1)}_{qf}}}. \label{Sp}
\eea

Using (\ref{Sp}) with one-particle distribution function  [$\eta_{f} = \tanh (\bar y_f/2)$], we express the effective fields~$\Delta_f$ through the order parameters $\eta_{f}$:  
\bea
&& \hspace{-2ex} \Delta_f  =  \frac{1}{2\beta} \ln \frac{1  +   \eta_{f}} {1-\eta_{f}} - \frac{1}{2}{\cal H}_f  - \frac{1}{2}{\boldsymbol{\muup}}_f {\bf E}.\nonumber
\eea
Then, $y_f$ are given by:
\bea
&& \hspace{-2ex} y_f  =  \frac{1}{2} \ln \frac{1  +   \eta_{f}} {1-\eta_{f}} + \frac{\beta}{2}{\cal H}_f  + \frac{\beta}{2}{\boldsymbol{\muup}}_f {\bf E}.\nonumber
\eea
From the first equality (\ref{Sp}), we obtain the system of equations for order parameters $\eta_{f}$:
\bea
&& \eta_{\frac{1}{3}}   =   \frac{1}{D}\Big(\sinh n_{1} \pm\sinh  n_{2}+a^{2}\sinh n_{3}\pm a^{2}\sinh  n_{4} \nonumber\\
&& ~~~~ + aa_{46}\sinh  n_{5}+\frac{a}{a_{46}}\sinh  n_{6} \mp aa_{46}\sinh  n_{7}\pm \frac{a}{a_{46}}\sinh  n_{8}\Big), \label{eta}\\
&& \eta_\frac{2}{4}  =  \frac{1}{D}\Big(\sinh  n_{1} \pm\sinh n_{2}- a^{2}\sinh  n_{3} \mp  a^{2}\sinh  n_{4} \nonumber\\ 
&& ~~~~ \mp aa_{46}\sinh  n_{5} \pm\frac{a}{a_{46}}\sinh  n_{6} +aa_{46}\sinh  n_{7}+ \frac{a}{a_{46}}\sinh  n_{8}\Big), 
\eea
where
\bea
&&   D = \cosh  n_{1} + \cosh  n_{2} + a^{2}\cosh  n_{3} +  a^{2}\cosh  n_{4}  \nonumber\\
&& ~~~~+aa_{46}\cosh  n_{5} + \frac{a}{a_{46}}\cosh  n_{6}
+ aa_{46}\cosh  n_{7} + \frac{a}{a_{46}}\cosh  n_{8}\,,\nonumber\\
&&a = \exp\bigg[-\beta \bigg( w^0 + \sum\limits_{l}\delta_i\varepsilon_i \bigg)\bigg], \qquad a_{46} = \exp[-\beta\left( \delta_4\varepsilon_4 + \delta_6\varepsilon_6 \right)],\nonumber\\
&&n_{1}=\frac{1}{2}(y_1 + y_2 + y_3 + y_4 ),~~~~  n_{2}=\frac{1}{2}(y_1 + y_2 - y_3 - y_4),\nonumber\\
&&n_{3}=\frac{1}{2}(y_1 - y_2 + y_3 - y_4), ~~~ n_{4}=\frac{1}{2}(y_1 - y_2 - y_3 + y_4),\nonumber\\
&&n_{5}=\frac{1}{2}(y_1 - y_2 + y_3 + y_4), ~~~ n_{6}=\frac{1}{2}(y_1 + y_2 + y_3 - y_4),\nonumber\\
&&n_{7}=\frac{1}{2}( -y_1 + y_2 + y_3 + y_4),~  n_{8}=\frac{1}{2}(y_1 + y_2 - y_3 + y_4).\nonumber
\eea

\section{Thermodynamic characteristics of GPI}

To calculate the dielectric, piezoelectric and elastic characteristics of the GPI, we use the thermodynamic potential per one primitive cell, obtained in the two-particle cluster approximation:
\bea
&& g  =  \frac{G}{N}  =  U_{\text{seed}} + H^0 -  2 \bigg(w^{0}  +  \sum\limits_{l} \delta_{l}\varepsilon_l\bigg) + 2k_{\text B}T\ln2 -
Nv\sum\limits_{j=1}^6\sigma_j \varepsilon_j \nonumber\\
&& ~~~ - \frac{1}{2}k_{\text B}T \sum\limits_{f=1}^4\ln \bigl( 1 - \eta_{f}^{2}\bigr)  - 2k_{\text B}T \ln D. \label{GG}
\eea

Minimizing the thermodynamic potential with respect to the strains $\varepsilon_{j}$, we have obtained a system of equations for the strains:
\bea
&&  \sigma_{l}=  c_{l1}^{E0}\varepsilon_1  +  c_{l2}^{E0}\varepsilon_2  +  c_{l3}^{E0}\varepsilon_3  +  c_{l5}^{E0}\varepsilon_5  -  e_{2l}^0E_2  -  \frac{2\delta_{l}}{v} +  \frac{2\delta_l}{v D}M_{\varepsilon}  \nonumber\\
&&~~~~ -  \frac{\psi_{11l}}{8v} (\eta_{1}^{2}+\eta_{3}^{2})-\frac{\psi_{13l}}{4v} \eta_{1}\eta_{3}-
\frac{\psi_{22l}}{8v} (\eta_{2}^{2}+\eta_{4}^{2})-   \frac{\psi_{24l}}{4v} \eta_{2}\eta_{4} \nonumber 
\eea
\bea
&&~~~~- \frac{\psi_{12l}}{4v} (\eta_{1}\eta_{2}+\eta_{3}\eta_{4})- \frac{\psi_{14l}}{4v} (\eta_{1}\eta_{4}+\eta_{2}\eta_{3}), \quad (l=1,2,3,5), \label{sigma}\\
&&  \sigma_{4} = c_{44}^{E0}\varepsilon_4+c_{46}^{E0}\varepsilon_6
- e_{14}^0 E_1 - e_{34}^0 E_3 + \frac{2\delta_4}{v D}M_{46} \nonumber\\
&&~~~~  -  \frac{\psi_{114}}{8v} (\eta_{1}^{2}-\eta_{3}^{2})-  \frac{\psi_{134}}{4v} \eta_{1}\eta_{3}-
\frac{\psi_{224}}{8v} (\eta_{2}^{2}-\eta_{4}^{2})-  \frac{\psi_{244}}{4v} \eta_{2}\eta_{4} \nonumber\\
&&~~~~- \frac{\psi_{124}}{4v} (\eta_{1}\eta_{2}-\eta_{3}\eta_{4})- \frac{\psi_{144}}{4v} (\eta_{1}\eta_{4}-\eta_{2}\eta_{3}),\\
&&  \sigma_{6} = c_{46}^{E0}\varepsilon_4 +
c_{66}^{E0}\varepsilon_6 - e_{16}^0 E_1 - e_{36}^0 E_3 + \frac{2\delta_6}{v D}M_{46} \nonumber\\
&&~~~~ -  \frac{\psi_{116}}{8v} (\eta_{1}^{2}-\eta_{3}^{2})-\frac{\psi_{136}}{4v} \eta_{1}\eta_{3}-
\frac{\psi_{226}}{8v} (\eta_{2}^{2}-\eta_{4}^{2})-    \frac{\psi_{246}}{4v} \eta_{2}\eta_{4} \nonumber\\
&&~~~~- \frac{\psi_{126}}{4v} (\eta_{1}\eta_{2}-\eta_{3}\eta_{4})- \frac{\psi_{146}}{4v} (\eta_{1}\eta_{4}-\eta_{2}\eta_{3}).
\eea
Here, the following notations are used:
\bea
&& M_{\varepsilon}  =  2a^{2}\cosh n_{3}+ 2a^{2}\cosh n_{4}
+aa_{46} \cosh n_{5}+ \frac{a}{a_{46}}\cosh n_{6}+aa_{46} \cosh n_{7} + \frac{a}{a_{46}}\cosh n_{8}\,,\nonumber\\
&& M_{46}=aa_{46} \cosh n_{5}- \frac{a}{a_{46}}\cosh n_{6}+aa_{46} \cosh n_{7} - \frac{a}{a_{46}}\cosh n_{8}.\nonumber
\eea

Differentiating the thermodynamic potential over fields $E_{i}$, we get the expressions for polarizations~$P_{i}$:
\bea
&& P_1   =  e_{14}^0\varepsilon_4   +  e_{16}^0\varepsilon_6  + \chi_{11}^{\varepsilon 0}E_1 +  \frac{1}{2v}\left[\mu_{13}^{x}(\eta_{1} - \eta_{3}) -  \mu_{24}^{x}(\eta_{2}  - \eta_{4})\right], \nonumber\\
&& P_2  =  e_{21}^0\varepsilon_1 + e_{22}^0\varepsilon_2 +
e_{23}^0\varepsilon_3 + e_{25}^0\varepsilon_5  + \chi_{22}^{\varepsilon 0}E_2  + \frac{1}{2v}\left[\mu_{13}^{y}(\eta_{1}+\eta_{3})-\mu_{24}^{y}(\eta_{2}+\eta_{4})\right],   \nonumber\\
&& P_3  =    e_{34}^0\varepsilon_4  +  e_{66}^0\varepsilon_6 +   \chi_{33}^{\varepsilon 0}E_3  +  \frac{1}{2v}\left[\mu_{13}^{z}(\eta_{1} - \eta_{3}) + \mu_{24}^{z}(\eta_{2} - \eta_{4})\right].\label{Pi}
\eea
The static isothermic dielectric susceptibilities of a mechanically clamped crystal
GPI are given by:
\bea
&&\chi_{11}^{\varepsilon}=\chi_{11}^{\varepsilon 0}+
\frac{1}{2v\Delta}\left[\mu_{13}^{x}
\big(\Delta_{1}^{\chi a}- \Delta_{3}^{\chi a}\big)-
\mu_{24}^{x}
\big(\Delta_{2}^{\chi a}- \Delta_{4}^{\chi a}\big)\right], \label{X11} \\
&&\chi_{22}^{\varepsilon}=\chi_{22}^{\varepsilon 0}+
\frac{1}{2v\Delta}\big[\mu_{13}^{y}
\big(\Delta_{1}^{\chi b}+ \Delta_{3}^{\chi b}\big)-
\mu_{24}^{y}
\big(\Delta_{2}^{\chi b}+ \Delta_{4}^{\chi b}\big)\big], \label{X22}
\\
&&\chi_{33}^{\varepsilon}=\chi_{33}^{\varepsilon 0}+
\frac{1}{2v\Delta}\left[\mu_{13}^{z}
\big(\Delta_{1}^{\chi c}- \Delta_{3}^{\chi c}\big)+
\mu_{24}^{z}
\big(\Delta_{2}^{\chi c}- \Delta_{4}^{\chi c}\big)\right]. \label{X33}
\eea
Here, the following notations are used:
\bea
&&  \hspace{-4ex} \Delta= \left| \begin{array}{cccc}
	2D  - \varkappa_{11} & -\varkappa_{12} & -\varkappa_{13} & -\varkappa_{14} \\
	-\varkappa_{21} & 2D  - \varkappa_{22} & -\varkappa_{23} & -\varkappa_{24} \\
	-\varkappa_{31} & -\varkappa_{32} & 2D  - \varkappa_{33} & -\varkappa_{34}\\
	-\varkappa_{41} & -\varkappa_{42} & -\varkappa_{43} & 2D  - \varkappa_{44}
\end{array}
\right|\nonumber,\\
&&   \hspace{-4ex}  \Delta_{1}^{\chi \alpha}= \left| \begin{array}{cccc}
	\varkappa_{1}^{\,\chi \alpha} & -\varkappa_{12} & -\varkappa_{13} & -\varkappa_{14} \vspace{0.8mm}\\
	\varkappa_{2}^{\,\chi \alpha} &   2D  - \varkappa_{22} & -\varkappa_{23} & -\varkappa_{24} \vspace{0.8mm}\\
	\varkappa_{3}^{\,\chi \alpha} & -\varkappa_{32} &   2D  - \varkappa_{33} & -\varkappa_{34}\vspace{0.8mm}\\
	\varkappa_{4}^{\,\chi \alpha} & -\varkappa_{42} & -\varkappa_{43} &   2D  - \varkappa_{44}
\end{array}
  \right|, ~~ \Delta_{3}^{\chi \alpha}= \left| \begin{array}{cccc}
	  2D  - \varkappa_{11} & -\varkappa_{12} & \varkappa_{1}^{\,\chi \alpha} & -\varkappa_{14} \vspace{0.8mm}\\
	-\varkappa_{21} &   2D  - \varkappa_{22} & \varkappa_{2}^{\,\chi \alpha} & -\varkappa_{24} \vspace{0.8mm}\\
	-\varkappa_{31} & -\varkappa_{32} & \varkappa_{3}^{\,\chi \alpha} & -\varkappa_{4}\vspace{0.8mm}\\
	-\varkappa_{41} & -\varkappa_{42} & \varkappa_{4}^{\,\chi \alpha}&   2D  - \varkappa_{44}
\end{array}
  \right|,\nonumber\\
&& \hspace{-4ex} \Delta_{2}^{\chi \alpha}= \left| \begin{array}{cccc}
	  2D  - \varkappa_{11} & \varkappa_{1}^{\,\chi \alpha} & -\varkappa_{13} & -\varkappa_{14} \vspace{0.8mm}\\
	-\varkappa_{21} & \varkappa_{2}^{\,\chi \alpha} & -\varkappa_{23} & -\varkappa_{24} \vspace{0.8mm}\\
	-\varkappa_{31} & \varkappa_{3}^{\,\chi \alpha} &   2D  - \varkappa_{33} & -\varkappa_{34}\vspace{0.8mm}\\
	-\varkappa_{41} & \varkappa_{4}^{\,\chi \alpha} & -\varkappa_{43} &   2D  - \varkappa_{44}
\end{array}
  \right|, ~~ \Delta_{4}^{\chi \alpha}= \left| \begin{array}{cccc}
	  2D  - \varkappa_{11} & -\varkappa_{12} & -\varkappa_{13} & \varkappa_{1}^{\,\chi \alpha} \vspace{0.8mm}\\
	-\varkappa_{21} &   2D  - \varkappa_{22} & -\varkappa_{23} & \varkappa_{2}^{\,\chi \alpha} \vspace{0.8mm}\\
	-\varkappa_{31} & -\varkappa_{32} &   2D  - \varkappa_{33} & \varkappa_{2}^{\,\chi \alpha}\vspace{0.8mm}\\
	-\varkappa_{41} & -\varkappa_{42} & -\varkappa_{43} & \varkappa_{4}^{\,\chi \alpha}
\end{array}
  \right|\nonumber,
\eea
where
\bea
&& \varkappa_{f1}= \varkappa_{f11}(\varphi_{1}^{+}+\beta\bar{\nu}_{1}^{+})+\varkappa_{f12}(\beta\nu_{2}^{+}+\beta\bar{\nu}_{2}^{+})
+\varkappa_{f13}(\varphi_{1}^{-}+\beta\bar{\nu}_{1}^{-})+\varkappa_{f14}\beta(\nu_{2}^{-}+\beta\bar{\nu}_{2}^{-}) , \nonumber\\
&& \varkappa_{f2}= \varkappa_{f12}(\varphi_{2}^{+}+\beta\bar{\nu}_{3}^{+})+\varkappa_{f11}(\beta\nu_{2}^{+}+\beta\bar{\nu}_{2}^{-})
+\varkappa_{f14}(\varphi_{2}^{-}+\beta\bar{\nu}_{3}^{-})+\varkappa_{f13}(\beta\nu_{2}^{-}+\beta\bar{\nu}_{2}^{+}), \nonumber\\
&& \varkappa_{f3}= \varkappa_{f11}(\varphi_{3}^{+}-\beta\bar{\nu}_{1}^{-})+\varkappa_{f12}(\beta\nu_{2}^{+}-\beta\bar{\nu}_{2}^{+})
-\varkappa_{f13}(\varphi_{3}^{-}-\beta\bar{\nu}_{1}^{+})-\varkappa_{f14}(\beta\nu_{2}^{-}-\beta\bar{\nu}_{2}^{-}), \nonumber\\
&& \varkappa_{f4}= \varkappa_{f12}(\varphi_{4}^{+}-\beta\bar{\nu}_{3}^{-})+\varkappa_{f11}(\beta\nu_{2}^{+}-\beta\bar{\nu}_{2}^{-})
-\varkappa_{f14}(\varphi_{4}^{-}-\beta\bar{\nu}_{3}^{+})-\varkappa_{f13}(\beta\nu_{2}^{-}-\beta\bar{\nu}_{2}^{+}), \nonumber\\
&&\varkappa_{f}^{\,\chi a}=\varkappa_{f13}\beta\mu_{13}^{x}+\varkappa_{f15}\beta\mu_{24}^{x}\,, \qquad \varkappa_{f}^{\,\chi b}= \varkappa_{f11}\beta\mu_{13}^{y}+\varkappa_{f12}\beta\mu_{24}^{y}\,, \qquad \varkappa_{f}^{\,\chi c}= \varkappa_{f13}\beta\mu_{13}^{z}+\varkappa_{f14}\beta\mu_{24}^{z}\,,\nonumber\\
&& \varphi_{1,3}^{\pm}=\frac{1}{1 -  \eta_{1,3}^{2}} +\beta\nu_{1}^{\pm}, \qquad \varphi_{2,4}^{\pm}=\frac{1}{1 -  \eta_{2,4}^{2}} +\beta\nu_{3}^{\pm}, \quad (f=1,2,3,4),\nonumber\\
&& \nu_{l}^{\pm}=\nu_{l}^{0\pm}+\left(\sum\limits_{i=1}^3\psi_{li}^{\pm}\varepsilon_{i}+
\psi_{l5}^{\pm}\varepsilon_{5}\right), \qquad \bar{\nu_{l}}^{\pm}=\psi_{l4}^{\pm}\varepsilon_{4}+\psi_{l6}^{\pm}\varepsilon_{6}\,,\nonumber\\
&& \nu_{1}^{0\pm}=\frac{1}{4}(J_{11}^{0}\pm J_{13}^{0});  \qquad \psi_{1i}^{\pm}=\frac{1}{4}(\psi_{11i}\pm\psi_{13i}),\nonumber\\
&& \nu_{2}^{0\pm}=\frac{1}{4}(J_{12}^{0}\pm J_{14}^{0}); \qquad \psi_{2i}^{\pm}=\frac{1}{4}(\psi_{12i}\pm\psi_{14i}),\nonumber\\
&& \nu_{3}^{0\pm}=\frac{1}{4}(J_{22}^{0}\pm J_{24}^{0}); \qquad \psi_{3i}^{\pm}=\frac{1}{4}(\psi_{22i}\pm\psi_{24i}),\nonumber\\
&& \varkappa_{\frac{1}{3}11}= (l_{1+3}^{c}+l_{5+6}^{c})-\eta_{\frac{1}{3}}(l_{1+3}^{s}+l_{5+6}^{s}), \qquad \varkappa_{\frac{1}{3}12}= (l_{1-3}^{c}\mp l_{7-8}^{c})-\eta_{\frac{1}{3}}(l_{1-3}^{s}+l_{7+8}^{s}),\nonumber\\
&&\varkappa_{\frac{1}{3}13}=\pm(l_{2+4}^{c}+l_{7+8}^{c})-\eta_{\frac{1}{3}}(l_{2+4}^{s}-l_{7-8}^{s}), \qquad \varkappa_{\frac{1}{3}14}=(\pm l_{2-4}^{c}-l_{5-6}^{c})-\eta_{\frac{1}{3}}(l_{2-4}^{s}-l_{5-6}^{s}),\nonumber\\
&&\varkappa_{\frac{2}{4}11}= (l_{1-3}^{c}\mp l_{5-6}^{c})-\eta_{\frac{2}{4}}(l_{1+3}^{s}+l_{5+6}^{s}), \qquad \varkappa_{\frac{2}{4}12}= (l_{1+3}^{c}+l_{7+8}^{c})-\eta_{\frac{2}{4}}(l_{1-3}^{s}+l_{7+8}^{s}),\nonumber\\
&&\varkappa_{\frac{2}{4}13}= (\pm l_{2-4}^{c}- l_{7-8}^{c})-\eta_{\frac{2}{4}}(l_{2+4}^{s}-l_{7-8}^{s}), \qquad \varkappa_{\frac{2}{4}14}= (\pm l_{2+4}^{c}\pm l_{5+6}^{c})-\eta_{\frac{2}{4}}(l_{2-4}^{s}-l_{5-6}^{s}),\nonumber\\
&&\varkappa_{\frac{1}{3}15}=(\mp l_{2-4}^{c}+l_{5-6}^{c})-\eta_{\frac{1}{3}}(-l_{2-4}^{s}+l_{5-6}^{s}), \qquad \varkappa_{\frac{2}{4}15}=\mp( l_{2+4}^{c}+l_{5+6}^{c})+\eta_{\frac{2}{4}}(-l_{2-4}^{s}+l_{5-6}^{s}),\nonumber\\
&& l_{1\pm3}^{c} =  \cosh n_{1} \pm  a^{2}\cosh n_{3}; \qquad  l_{2\pm4}^{c} =  \cosh n_{2} \pm  a^{2}\cosh n_{4};\nonumber\\
&&~l_{5\pm6}^{c} =  aa_{46}\cosh n_{5} \pm  \frac{a}{a_{46}}\cosh n_{6}; \qquad l_{7\pm8}^{c} =  aa_{46}\cosh n_{7} \pm  \frac{a}{a_{46}}\cosh n_{8};\nonumber\\
&&l_{1\pm3}^{s} =  \sinh n_{1} \pm  a^{2}\sinh n_{3}; \qquad  l_{2\pm4}^{s} =  \sinh n_{2} \pm  a^{2}\sinh n_{4};\nonumber\\
&&~l_{5\pm6}^{s} =  aa_{46}\sinh n_{5} \pm  \frac{a}{a_{46}}\sinh n_{6}; \qquad l_{7\pm8}^{s} =  aa_{46}\sinh n_{7} \pm  \frac{a}{a_{46}}\sinh n_{8}.\nonumber
\eea
Differentiating  the expressions (\ref{Pi}) over strains $\varepsilon_i$ at a constant field, we obtain the expressions for isothermic coefficients of piezoelectric stress  $e_{2l}$ ($l=1, 2, 3, 5$):
\bea
&& e_{2l}^{} =   \left( \frac{\partial P_2}{\partial
	\varepsilon_l}\right)_{E_2}   = e_{2l}^0+ \frac{\mu_{13}^{y}}{2v\Delta}(\Delta_{1l}^{e}+\Delta_{3l}^{e})
-\frac{\mu_{24}^{y}}{2v\Delta}(\Delta_{2l}^{e}+\Delta_{4l}^{e}), \quad (l=1,2,3,5),
\eea
where the following notations are used:
\bea
&&   \hspace{-4ex}  \Delta_{1l}^{e}= \left| \begin{array}{cccc}
	\varkappa_{1l}^{e} & -\varkappa_{12} & -\varkappa_{13} & -\varkappa_{14} \vspace{0.7mm}\\
	\varkappa_{2l}^{e} &   2D  - \varkappa_{22} & -\varkappa_{23} & -\varkappa_{24} \vspace{0.7mm}\\
	\varkappa_{3l}^{e} & -\varkappa_{32} &   2D  - \varkappa_{33} & -\varkappa_{34}\vspace{0.7mm}\\
	\varkappa_{4l}^{e} & -\varkappa_{42} & -\varkappa_{43} &   2D  - \varkappa_{44}
\end{array}
  \right|,~~~\Delta_{3l}^{e}= \left| \begin{array}{cccc}
	  2D  - \varkappa_{11} & -\varkappa_{12} & \varkappa_{1l}^{e} & -\varkappa_{14} \vspace{0.7mm}\\
	-\varkappa_{21} &   2D  - \varkappa_{22} & \varkappa_{2l}^{e} & -\varkappa_{24} \vspace{0.7mm}\\
	-\varkappa_{31} & -\varkappa_{32} & \varkappa_{3l}^{e} & -\varkappa_{4}\vspace{0.7mm}\\
	-\varkappa_{41} & -\varkappa_{42} & \varkappa_{4l}^{e}&   2D  - \varkappa_{44}
\end{array}
  \right|,\nonumber\\
&& \hspace{-4ex} \Delta_{2l}^{e}= \left| \begin{array}{cccc}
	  2D  - \varkappa_{11} & \varkappa_{1l}^{e} & -\varkappa_{13} & -\varkappa_{14} \vspace{0.7mm}\\
	-\varkappa_{21} & \varkappa_{2l}^{e} & -\varkappa_{23} & -\varkappa_{24} \vspace{0.7mm}\\
	-\varkappa_{31} & \varkappa_{3l}^{e} &   2D  - \varkappa_{33} & -\varkappa_{34}\vspace{0.7mm}\\
	-\varkappa_{41} & \varkappa_{4l}^{e} & -\varkappa_{43} &   2D  - \varkappa_{44}
\end{array}
  \right|,~~~\Delta_{4l}^{e}= \left| \begin{array}{cccc}
	  2D  - \varkappa_{11} & -\varkappa_{12} & -\varkappa_{13} & \varkappa_{1l}^{e} \vspace{0.7mm}\\
	-\varkappa_{21} &   2D  - \varkappa_{22} & -\varkappa_{23} & \varkappa_{2l}^{e} \vspace{0.7mm}\\
	-\varkappa_{31} & -\varkappa_{32} &   2D  - \varkappa_{33} & \varkappa_{2l}^{e}\vspace{0.7mm}\\
	-\varkappa_{41} & -\varkappa_{42} & -\varkappa_{43} & \varkappa_{4l}^{e}
\end{array}
  \right|\nonumber,
\eea
\bea
&&\varkappa_{fl}^{e}=\beta(\psi_{1l}^{+}\varkappa_{f11}+\psi_{2l}^{+}\varkappa_{f12})(\eta_{1}+\eta_{3}) +
\beta(\psi_{2l}^{+}\varkappa_{f11}+\psi_{3l}^{+}\varkappa_{f12})(\eta_{2}+\eta_{4})\nonumber\\
&&~~~~~+\beta(\psi_{1l}^{-}\varkappa_{f13}+\psi_{2l}^{-}\varkappa_{f14})(\eta_{1}-\eta_{3})+
\beta(\psi_{2l}^{-}\varkappa_{f13}+\psi_{3l}^{-}\varkappa_{f14})(\eta_{2}-\eta_{4})+2\beta\delta_{l}(\rho_{f1}+\rho_{f2}),\nonumber\\
&&\psi_{1l}^{\pm}=\frac{1}{4}(\psi_{11l}\pm\psi_{13l}), \qquad \psi_{2l}^{\pm}=\frac{1}{4}(\psi_{12l}\pm\psi_{14l}),
 \qquad \psi_{3l}^{\pm}=\frac{1}{4}(\psi_{22l}\pm\psi_{24l}),
\nonumber\\
&&\rho_{\frac{1}{3}1}=-2(l^{s}_{3\pm4}-\eta_{\frac{1}{3}}l^{c}_{3+4}), \qquad  \rho_{\frac{1}{3}2}=-l^{s}_{5+6}\pm l^{s}_{7-8}+\eta_{\frac{1}{3}}(l^{c}_{5+6}+l^{c}_{7+8}),\nonumber\\
&&\rho_{\frac{2}{4}1}=2(l^{s}_{3\pm4}+\eta_{\frac{1}{3}}l^{c}_{3+4}),  \qquad \rho_{\frac{2}{4}2}=\pm l^{s}_{5-6}- l^{s}_{7+8}+\eta_{\frac{1}{3}}(l^{c}_{5+6}+l^{c}_{7+8}),\nonumber\\
&&\rho_{\frac{1}{3}j}=l^{s}_{5-6}\mp l^{s}_{7+8}-\eta_{\frac{1}{3}}(l^{c}_{5-6}+l^{c}_{7-8}),  \qquad 
\rho_{\frac{2}{4}j}=\mp l^{s}_{5+6}+ l^{s}_{7-8}-\eta_{\frac{2}{4}}(l^{c}_{5-6}+l^{c}_{7-8}),\nonumber\\
&&l_{3\pm4}^{s}=  a^{2}\sinh n_{3}\pm a^{2}\sinh n_{4}\,, \qquad l_{3+4}^{c}=  a^{2}\cosh n_{3}+ a^{2}\cosh n_{4}. \nonumber
\eea
Constants of piezoelectric stress are obtained by differentiating the electric field, found from (\ref{sigma}), over the strains at a constant polarization:
\be
h_{2l} = -\left( \frac{\partial E_2}{\partial \varepsilon_l} \right)_{P_2} = \frac{e_{2l}}{\chi_{22}^{\varepsilon}}. 
\ee

Molar entropy of the proton subsystem
(here, \textit{R} is the gas constant):
\bea
&&\hspace{-6ex} S = - \frac{R}{4} \left( \frac{\partial g}{\partial T} \right)_{\eta,\varepsilon_i} = \frac{R}{4}
\bigg( - 2\ln2 + \sum\limits_{f=1}^4\ln\bigl( 1 - \eta_{f} \bigr)+ 2\ln D   \nonumber\\
&&\hspace{-6ex} ~~- 2 \big\{\beta\nu_{1}^{+}(\eta_{1}+\eta_{3})^{2}+\beta\bar{\nu}_{1}^{+}[\eta_{1}(\eta_{1}+\eta_{3})+ \eta_{3}(\eta_{1}-\eta_{3})] + 2\beta\nu_{2}^{+}(\eta_{1}+\eta_{3})(\eta_{2}+\eta_{4})\nonumber\\ 
&&\hspace{-6ex} ~~+2\beta\bar{\nu}_{2}^{+}(\eta_{1}-\eta_{3})(\eta_{2}+\eta_{4})+\beta\nu_{3}^{+}(\eta_{2}+\eta_{4})^{2}+\beta\bar{\nu}_{3}^{+}
[\eta_{2}(\eta_{2}+\eta_{4})+ \eta_{4}(\eta_{2}-\eta_{4})]\nonumber\\
&&\hspace{-6ex} ~~+\beta\nu_{1}^{-}(\eta_{1}-\eta_{3})^{2}+ \beta\bar{\nu}_{1}^{-}[\eta_{1}(\eta_{1}-\eta_{3})+ \eta_{3}(\eta_{1}+\eta_{3})]+2\beta\nu_{2}^{-}(\eta_{1}-\eta_{3})(\eta_{2}-\eta_{4})\nonumber\\
&&\hspace{-6ex} ~~+2\beta\bar{\nu}_{2}^{-}(\eta_{1}+\eta_{3})(\eta_{2}-\eta_{4})+ \beta\nu_{3}^{-}(\eta_{2}-\eta_{4})^{2}+ \beta\bar{\nu}_{3}^{-}[\eta_{2}(\eta_{2}-\eta_{4})- \eta_{4}(\eta_{2}+\eta_{4})]\big\}+\frac{4w}{T D} M \bigg). \label{S}
\eea

The molar heat capacity of the proton subsystem of
GPI crystals can be found from the entropy (\ref{S}):
\bea
&&\Delta C^{\sigma} = T \left( \frac{\partial S}{\partial T} \right)_{\sigma}.
\eea

\section{Comparison of theoretical results with the experimental data}

\setcounter{equation}{0}
\renewcommand{\theequation}{4.\arabic{equation}}

To calculate the temperature and field dependences of
dielectric and piezoelectric characteristics
of GPI, we have to determine the values of the following model parameters:
\begin{itemize}
\item parameter of short-range interactions $w^{0}$;
\item parameters of long-range interactions $\nu_{f}^{0\pm}$ ($f=1,2,3$);
\item deformational potentials $\delta_{i}$,   $\psi_{fi}^{\pm}$ ($f=1,2,3$; $i=1,\ldots,6$);
\item components of effective dipole moments
$\mu_{13}^{x}$; $\mu_{24}^{x}$; $\mu_{13}^{y}$; $\mu_{24}^{y}$; $\mu_{13}^{z}$; $\mu_{24}^{z}$;
\item ``seed'' dielectric susceptibilities $\chi_{ij}^{\varepsilon 0}$;
\item ``seed'' coefficients of piezoelectric stress $e_{ij}^0$;
\item ``seed''  elastic constants $c_{ij}^{E0}$.
\end{itemize}

To determine the above listed parameters we use the measured temperature dependences  for the set of physical
characteristics of GPI,  namely  $P_s(T)$ \cite{nay1}, $ C_p(T)$ \cite{shi3}, $\varepsilon_{11}^\sigma$,  $\varepsilon_{33}^\sigma$ \cite{dac}, $d_{21}$, $d_{23}$ \cite{wie},  as well as the dependence of phase transition temperature $T_{\text c}(p)$ \cite{yas,yas1}  on hydrostatic pressure.

The volume of primitive cell of GPI is the $v = 0.601\cdot 10^{-21}$~cm$^3$.

Numerical analysis shows that thermodynamic characteristics depend on two linear combinations of long-range interactions
$\nu^{0+}=\nu_1^{0+}+2\nu_2^{0+}+\nu_3^{0+}$ and $\nu^{0-}=\nu_1^{0-}+2\nu_2^{0-}+\nu_3^{0-}$, and practically do not depend  (deviation $<0.1$\%) on  separate values of the $\nu_{f}^{0\pm}$ at given $\nu^{0+}$ and $\nu^{0-}$. The optimal values of these combinations are  $\nu^{0+}/k_{\text B}=10.57$~K, $\nu^{0-}/k_{\text B}=-0.8$~K; and as concrete values of the  $\nu_{f}^{0\pm}$ we use $\tilde \nu_1^{0+}=\tilde \nu_2^{0+}=\tilde \nu_3^{0+}=2.643$~K, $\tilde \nu_1^{0-}=\tilde \nu_2^{0-}=\tilde \nu_3^{0-}=0.2$~K, where $\tilde \nu_{f}^{0\pm}=\nu_{f}^{0\pm}/k_{\text B}$. 

The calculated parameter of  short-range interactions $w^0(x)$ of GPI crystal is equal to $w^0/k_{\text B}=800$~K.
The optimal values of  deformational potentials $\delta_{i}$:
$\tilde\delta_{1}=500$~K,  $\tilde\delta_{2}=600$~K, $\tilde\delta_{3}=500$~K, $\tilde\delta_{4}=150$~K, $\tilde\delta_{5}=100$~K, $\tilde\delta_{6}=150$~K; $\tilde\delta_{i}={\delta_{i}}/{k_{\text B}}$.

The optimal values of the $\psi_{fi}^{\pm}$ are as follows:
$\tilde\psi_{f1}^{+} = 93.6$~K,  $\tilde\psi_{f2}^{+} = 252.5$~K,  $\tilde\psi_{f3}^{+} = 110.7$~K,
$\tilde\psi_{f4}^{+} = \tilde\psi_{f6}^{+} = \tilde\psi_{f4}^{-}=\tilde\psi_{f6}^{-}=79.5$~K,  $\tilde\psi_{f5}^{+} = 22.7$~K,  $\tilde\psi_{f1}^{-}=\tilde\psi_{f2}^{-}=\tilde\psi_{f3}^{-}=\tilde\psi_{f5}^{-}=0$~K, 
where
$\tilde\psi_{fi}^{\pm} =\psi_{fi}^{\pm}/{k_{\text B}}$.

The effective dipole moments in the paraelectric phase are equal to $\mu_{13}^{x}=0.4\cdot 10^{-18}$~esu$\cdot$cm; $\mu_{13}^{y}=4.05\cdot 10^{-18}$~esu$\cdot$cm; $\mu_{13}^{z}=4.2\cdot 10^{-18}$~esu$\cdot$cm; $\mu_{24}^{x}=2.3\cdot 10^{-18}$~esu$\cdot$cm; $\mu_{24}^{y}=3.0\cdot 10^{-18}$~esu$\cdot$cm; $\mu_{24}^{z}=2.2\cdot 10^{-18}$~esu$\cdot$cm. 
In the ferroelectric phase, the $y$-component of the first dipole moment is  $\mu_{13\text{ferro}}^{y}=3.82\cdot 10^{-18}$~esu$\cdot$cm.
A larger dipole moment ${\boldsymbol{\muup}}_{13}$ in comparison with ${\boldsymbol{\muup}}_{24}$, and different values of  $\mu_{13}^{y}$ in the paraelectric and ferroelectric phases agree with the results of lattice dynamics simulations \cite {Shchur2014}, where vibrations of the oxygen atoms connected with proton 1 (in our notations), are much more intensive than the vibrations of oxygen atoms revealed near proton 2.

\begin{figure}[!b]
\centering
\includegraphics[scale=0.9]{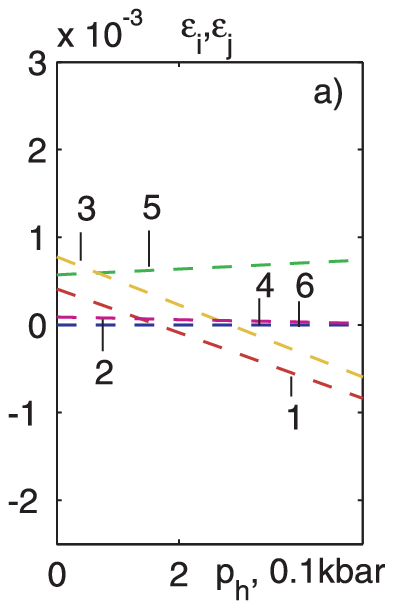}\includegraphics[scale=0.9]{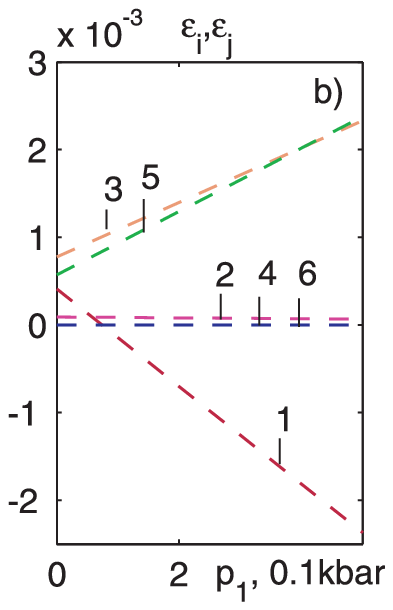}\includegraphics[scale=0.9]{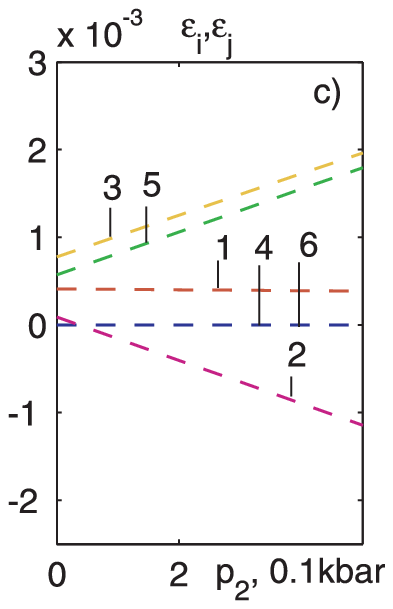}\includegraphics[scale=0.9]{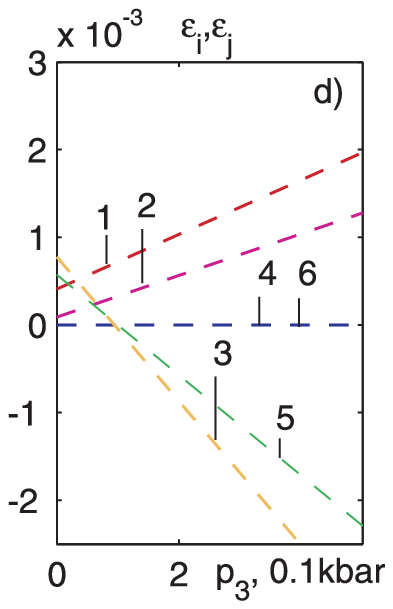}
\caption{(Colour online) The dependences of strains $\varepsilon_{1}$ (curves 1), $\varepsilon_{2}$ (2), $\varepsilon_{3}$ (3), $\varepsilon_{4}$ (4), $\varepsilon_{5}$ (5), $\varepsilon_{6}$ (6) on hydrostatic pressure  $p_{\text h}$ (figure a) and uniaxial pressures $p_{1}$ (b), $p_{2}$ (c), $p_{3}$ (d)  at  $\Delta T = -5$~K.} \label{eiej_h1}
\end{figure}
\begin{figure}[!t]
	\centering
		\includegraphics[scale=0.9]{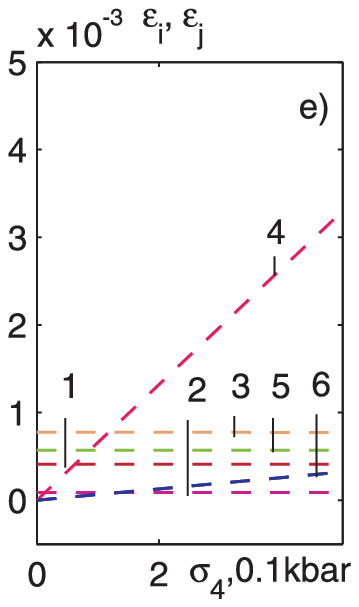}\quad\includegraphics[scale=0.9]{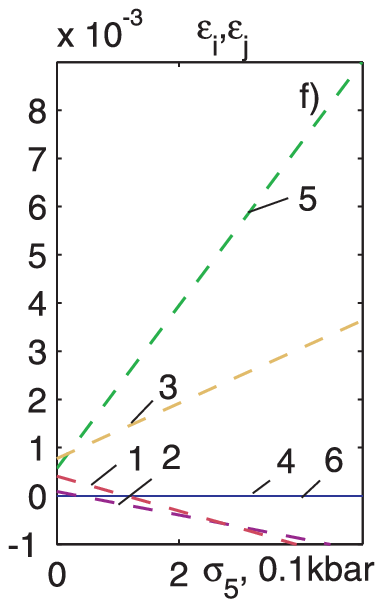}\quad\includegraphics[scale=0.9]{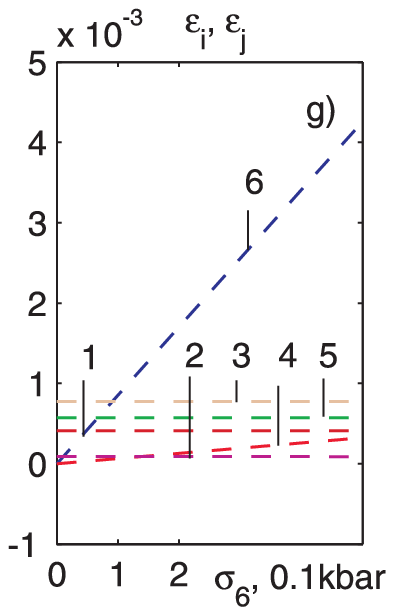}
	\caption[]{(Colour online) The dependences of strains $\varepsilon_{1}$ (curves 1), $\varepsilon_{2}$ (2), $\varepsilon_{3}$ (3), $\varepsilon_{4}$ (4), $\varepsilon_{5}$ (5), $\varepsilon_{6}$ (6) on shear stresses  $\sigma_{4}$(e), $\sigma_{5}$(f),  $\sigma_{6}$(g)  at $\Delta T = -5$~K.} \label{eiej_sigma4}
\end{figure}

``Seed'' coefficients of piezoelectric stress, dielectric susceptibilities and elastic constants\\
$e_{ij}^0 = 0.0~\frac{\text{esu}}{\text{cm}^2}$; \quad 
~$\chi_{11}^{\varepsilon 0} = 0.1$, \quad  $\chi_{22}^{\varepsilon 0}= 0.403$, \quad  $\chi_{33}^{\varepsilon 0} = 0.5$, \quad  $\chi_{13}^{\varepsilon 0} = 0.0$;\\
$c_{11}^{0E}   =   26.91\cdot10^{10}~\frac{\text{dyn}}{\text{cm}^2}$\,, \quad 
$c_{12}^{E0}   =   14.5 \cdot 10^{10}~\frac{\text{dyn}}{\text{cm}^2}$\,, \quad 
$c_{13}^{E0}   =   11.64 \cdot10^{10}~\frac{\text{dyn}}{\text{cm}^2}$\,, \quad 
$c_{15}^{E0} = 3.91  \cdot10^{10}~\frac{\text{dyn}}{\text{cm}^2}$\,,\\
$c_{22}^{E0} = [64.99 -  0.04(T-T_{\text c})] \cdot10^{10}~\frac{\text{dyn}}{\text{cm}^2}$\,, \quad 
$c_{23}^{E0} = 20.38\cdot10^{10}~\frac{\text{dyn}}{\text{cm}^2}$\,, \quad 
$c_{25}^{E0} = 5.64  \cdot10^{10}~\frac{\text{dyn}}{\text{cm}^2}$\,, \\
$c_{33}^{E0} = 24.41\cdot10^{10}~\frac{\text{dyn}}{\text{cm}^2}$\,, \quad 
$c_{35}^{E0} = -2.84  \cdot10^{10}~\frac{\text{dyn}}{\text{cm}^2}$\,, \quad 
$c_{55}^{E0} = 8.54 \cdot 10^{10}~\frac{\text{dyn}}{\text{cm}^2}$\,,\\
$c_{44}^{E0} = 15.31 \cdot10^{10}~\frac{\text{dyn}}{\text{cm}^2}$\,, \quad 
$c_{46}^{E0} = -1.1 \cdot 10^{10}~\frac{\text{dyn}}{\text{cm}^2}$\,, \quad 
$c_{66}^{E0} = 11.88 \cdot 10^{10}~\frac{\text{dyn}}{\text{cm}^2}$.

Now, let us focus on the obtained results.
The influence of hydrostatic pressure $p_{\text h}$, uniaxial pressures  $p_{1}$, $p_{2}$, $p_{3}$ and mechanical shear stresses  $\sigma_{4}$, $\sigma_{5}$, $\sigma_{6}$ first of all causes the changes in the strains $ \varepsilon_{j}$ of the crystal.
In figure~\ref{eiej_h1} there are presented the dependences of strains $\varepsilon_j$ of GPI  crystal on the hydrostatic $p_{\text h}$ and on uniaxial pressures $p_i$, and in figure~\ref{eiej_sigma4} --- on the shear stresses $\sigma_{j}$ at temperature difference  $\Delta T = T-T_{\text c} = -5$~K.

\begin{figure}[!t]
	\centering
		\includegraphics[scale=0.9]{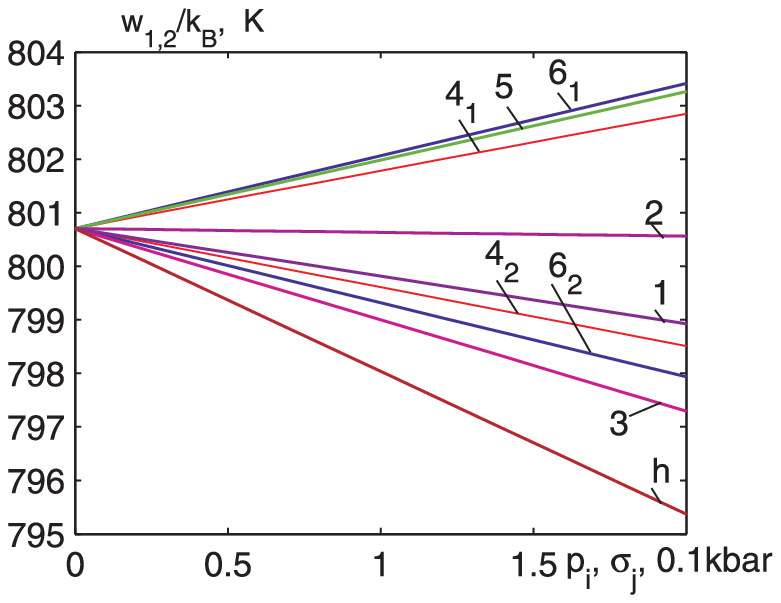}~~~~\includegraphics[scale=0.9]{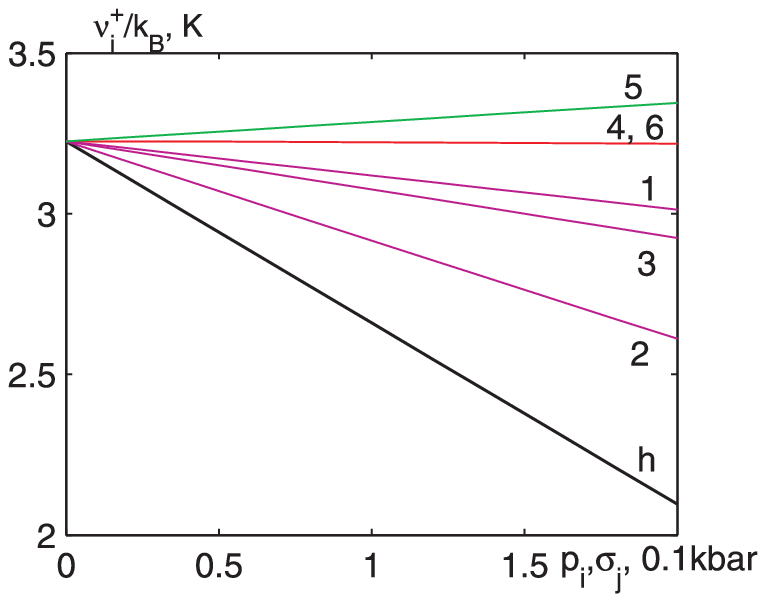} 
	\caption[]{(Colour online) Dependences of parameters of short-range interactions $w_{1,2}$ and  long-range interactions $\nu^+_{1, 2,3}$ of GPI crystal on hydrostatic pressure (curve~h), uniaxial pressures $p_{1}$ (1), $p_{2}$ (2), $p_{3}$~(3) and shear stresses    $\sigma_{4}$ (4), $\sigma_{5}$ (5), $\sigma_{6}$ (6) at  $\Delta T=-5$~K.  } \label{w12_p_sigma}
\end{figure}
\begin{figure}[!t]
\centering
		\includegraphics[scale=0.95]{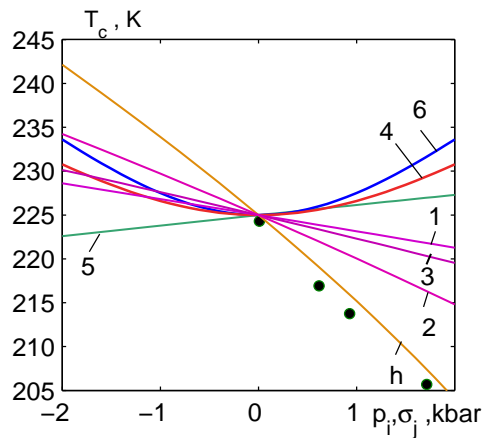}
	\caption[]{(Colour online) Dependences of the phase transition temperature  $T_{\text c}$  of GPI crystal on hydrostatic pressure (curve~h), uniaxial pressures $p_{1}$ (1), $p_{2}$ (2), $p_{3}$ (3) and shear stresses    $\sigma_{4}$ (4), $\sigma_{5}$ (5), $\sigma_{6}$ (6). Symbols $\bullet$ are experimental data \cite{yas}.} \label{Tc_h123456}
\end{figure}

The hydrostatic pressure $p_{\text h}$, uniaxial pressures $p_{1}$, $p_{2}$, $p_{3}$ and shear stress $\sigma_{5}$ do not change the symmetry of the crystal. Therefore, the strains $\varepsilon_4=0$ and $\varepsilon_6=0$. Other strains almost linearly change with pressure.
In particular, the pressure $p_{\text h}$ causes a decrease of the strains $\varepsilon_1$ and $\varepsilon_3$ and a weaker decrease of the strain $\varepsilon_2$, while the strain $\varepsilon_5$ weakly increases.
When the pressure  $p_{1}$ increases, then the strain~$\varepsilon_1$ decreases, while the strain $\varepsilon_2$ decreases much weaker, but the strains $\varepsilon_3$ and $\varepsilon_5$ increase.
In the presence of uniaxial pressure $p_{2}$,  the strain $\varepsilon_2$ decreases,  the strain $\varepsilon_1$ decreases much weaker, but  the strains $\varepsilon_3$ and $\varepsilon_5$ increase.
In the case of pressure $p_{3}$,  the strains $\varepsilon_3$  and $\varepsilon_5$ decrease, but  the strains $\varepsilon_1$ and $\varepsilon_2$ increase.
Application of the shear stress $\sigma_{5}$ causes an increase of the strains $\varepsilon_5$, $\varepsilon_3$, and a decrease of the  strains~$\varepsilon_1$ and $\varepsilon_2$.

\begin{figure}[!b]
\centering
\begin{minipage}{0.49\textwidth}
\begin{center}
\includegraphics[width=0.95\textwidth]{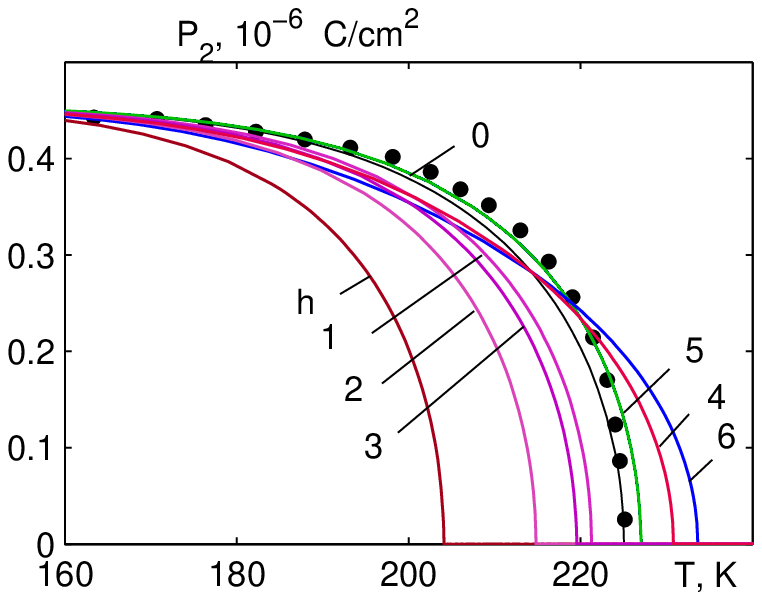}
\caption{(Colour online) The temperature dependences of polarization $P_{2}$  of GPI crystal at zero stresses (curve 0), under hydrostatic pressure  (curve~h), uniaxial pressures $p_{1}$ (1), $p_{2}$ (2), $p_{3}$ (3) and different shear stresses  $\sigma_{4}$ (4), $\sigma_{5}$ (5), $\sigma_{6}$ (6). Value of the pressures and stresses is  2~kbar. Symbols $\bullet$ are experimental data \cite{nay1}.}  \label{P2_h123456}
\end{center}
\end{minipage}
\begin{minipage}{0.49\textwidth}
\begin{center}
\includegraphics[width=0.99\textwidth]{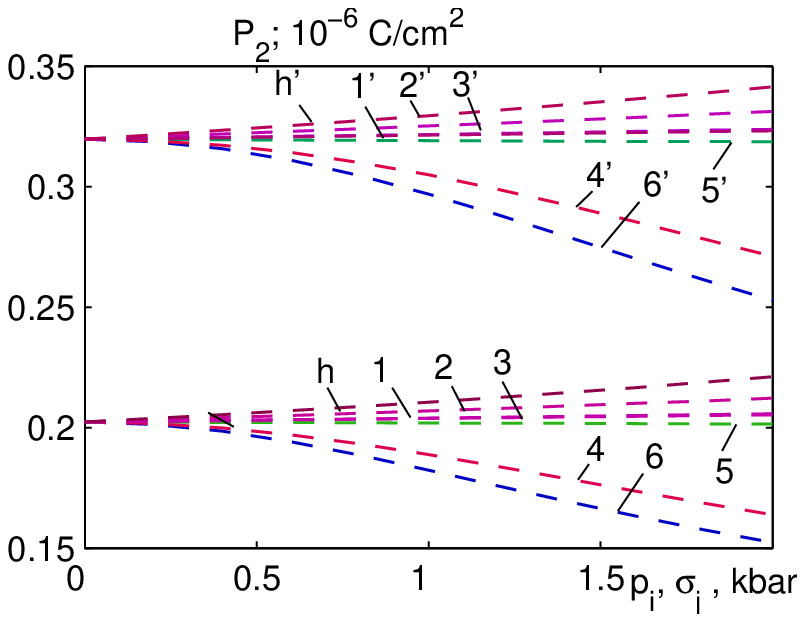}
\caption{(Colour online) Dependences of polarization $P_{2}$ of GPI crystal on hydrostatic pressure  (curve~h), uniaxial pressures ($p_{1}$~---~1, $p_{2}$~---~2, $p_{3}$~---~3) and shear stresses  ($\sigma_{4}$~---~4, $\sigma_{5}$~---~5, $\sigma_{6}$~---~6) at different  $\Delta T$: $-5$~K~---~h,1,2,3,4,5,6; $-15$~K~---~h$'$,1$'$,2$'$,3$'$,4$'$,5$'$,6$'$. } \label{P2_p_sigma}
\end{center}
\end{minipage}
\end{figure}
\begin{figure}[!b]
\centering
\begin{minipage}{0.49\textwidth}
\begin{center}
\includegraphics[width=0.95\textwidth]{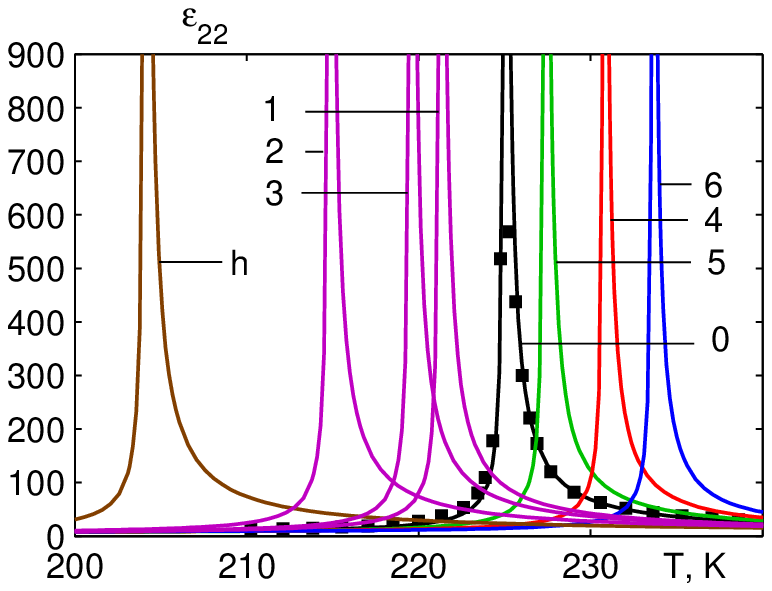}
\caption{(Colour online) The temperature dependences of dielectric permittivity  $\varepsilon_{22}$ of GPI crystal   at zero stresses (curve~0),  under hydrostatic pressure (h), uniaxial pressures ($p_{1}$~---~1, $p_{2}$~---~2, $p_{3}$~---~3) and shear stresses   ($\sigma_{4}$~---~4, $\sigma_{5}$~---~5, $\sigma_{6}$~---~6). Value of the pressures and stresses is  2~kbar. Symbols $\blacksquare$ are experimental data taken from \cite{wie}.} \label{eps22_h123456}
\end{center}
\end{minipage}
\begin{minipage}{0.49\textwidth}
\begin{center}
\includegraphics[width=0.95\textwidth]{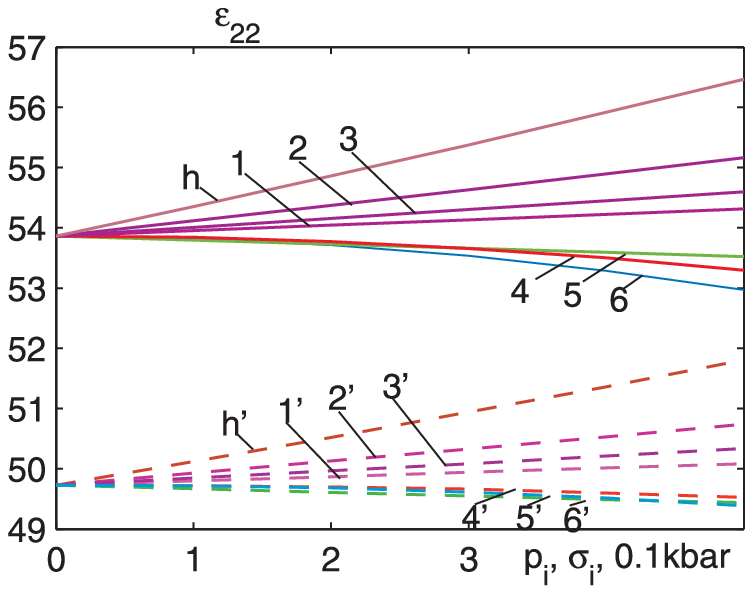}
\caption{(Colour online) Dependences of dielectric permittivity  $\varepsilon_{22}$ of GPI crystal on hydrostatic pressure  (curve~h), uniaxial pressures ($p_{1}$~---~1, $p_{2}$~---~2, $p_{3}$~---~3) and shear stresses  ($\sigma_{4}$~---~4, $\sigma_{5}$~---~5, $\sigma_{6}$~---~6) at different  $\Delta T$: 6.5~K~---~h,1,2,3,4,5,6; $-2.0$~K~---~h$'$,1$'$,2$'$,3$'$,4$'$,5$'$,6$'$. } \label{eps22_p_sigma}
\end{center}
\end{minipage}
\end{figure}

The influence of the shear stress $\sigma_{4}$ and $\sigma_{6}$  causes an increase of the strains   $\varepsilon_4$ and $\varepsilon_6$, but other strains practically do not change.

The change of strains leads to the changes in parameters of short-range interactions $w_{1}$, $w_{2}$
and long-range interactions $\nu^+_{1, 2,3}$ (figure~\ref{w12_p_sigma}), which leads to changes in the mean values of pseudospins  $\eta_{f}$ and in all thermodynamic characteristics.

\begin{figure}[!b]
\centering
\begin{minipage}{0.49\textwidth}
\begin{center}
\includegraphics[width=0.95\textwidth]{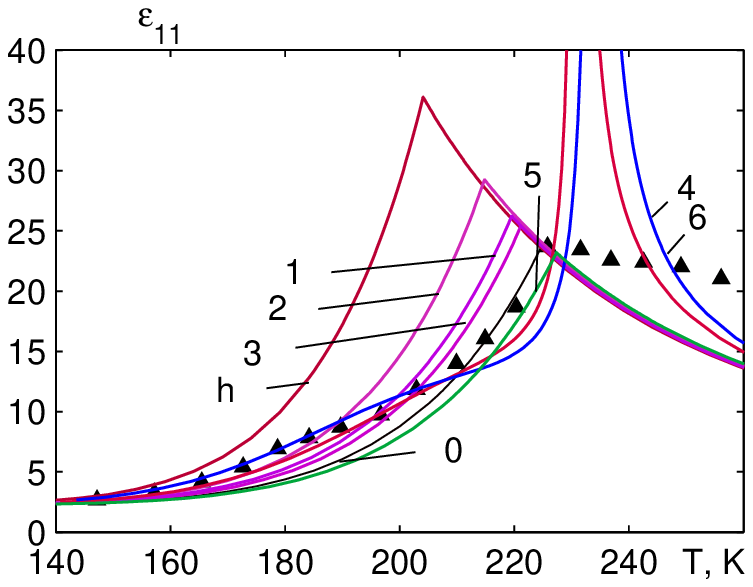}
\caption{(Colour online) The temperature dependences of dielectric permittivity $\varepsilon_{11}$ of GPI crystal   at zero stresses (curve~0),  under hydrostatic pressure (h), uniaxial pressures ($p_{1}$~---~1, $p_{2}$~---~2, $p_{3}$~---~3) and shear stresses   ($\sigma_{4}$~---~4, $\sigma_{5}$~---~5, $\sigma_{6}$~---~6). Value of the pressures and stresses is  2~kbar. Symbols $ \blacktriangle$ are experimental data taken  from \cite{dac}.} \label{eps11_h123456}
\end{center}
\end{minipage}
\begin{minipage}{0.49\textwidth}
\begin{center}
\vspace{-3mm}
\includegraphics[width=0.95\textwidth]{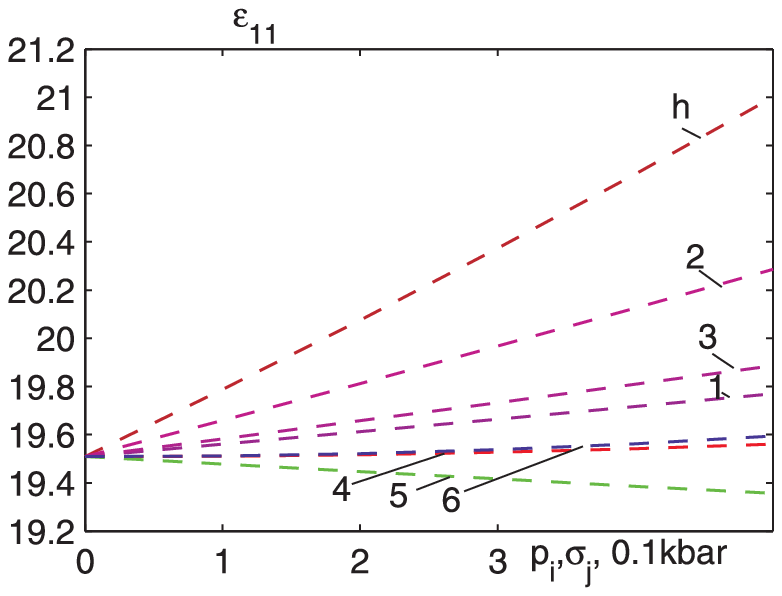}
\caption{(Colour online) Dependences of dielectric permittivity $\varepsilon_{11}$ of GPI crystal on hydrostatic pressure  (curve~h), uniaxial pressures ($p_{1}$~---~1, $p_{2}$~---~2, $p_{3}$~---~3) and shear stresses  ($\sigma_{4}$~---~4, $\sigma_{5}$~---~5, $\sigma_{6}$~---~6) at $\Delta T=-5$~K.} \label{eps11_p_sigma}
\end{center}
\end{minipage}
\end{figure}

\begin{figure}[!b]
\centering
\begin{minipage}{0.49\textwidth}
\centering
\includegraphics[width=0.95\textwidth]{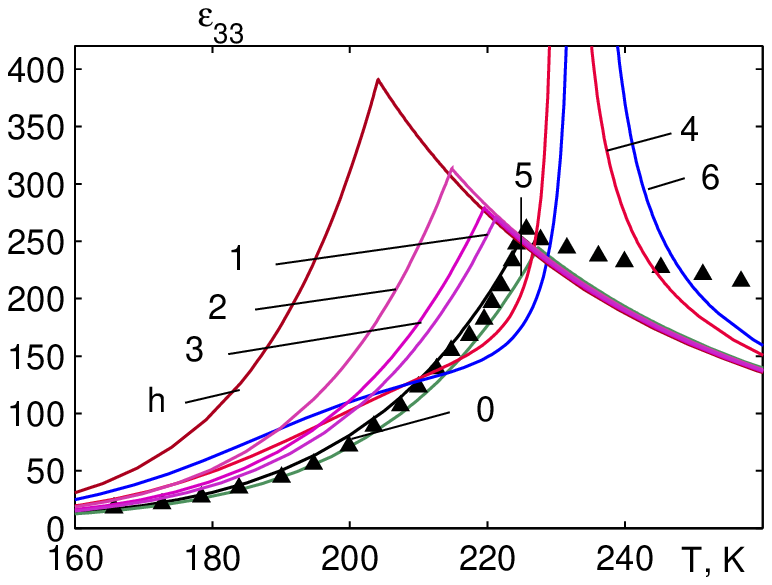}
\caption{(Colour online) The temperature dependences of dielectric permittivity $\varepsilon_{33}$ of GPI crystal   at zero stresses (curve~0),  under hydrostatic pressure (h), uniaxial pressures ($p_{1}$~---~1, $p_{2}$~---~2, $p_{3}$~---~3) and shear stresses   ($\sigma_{4}$~---~4, $\sigma_{5}$~---~5, $\sigma_{6}$~---~6). Value of the pressures and stresses is  2~kbar. Symbols $ \blacktriangle$ are experimental data taken  from \cite{dac}.} \label{eps33_h123456}
\end{minipage}
\begin{minipage}{0.49\textwidth}
\centering
\vspace{-3mm}
\includegraphics[width=0.95\textwidth]{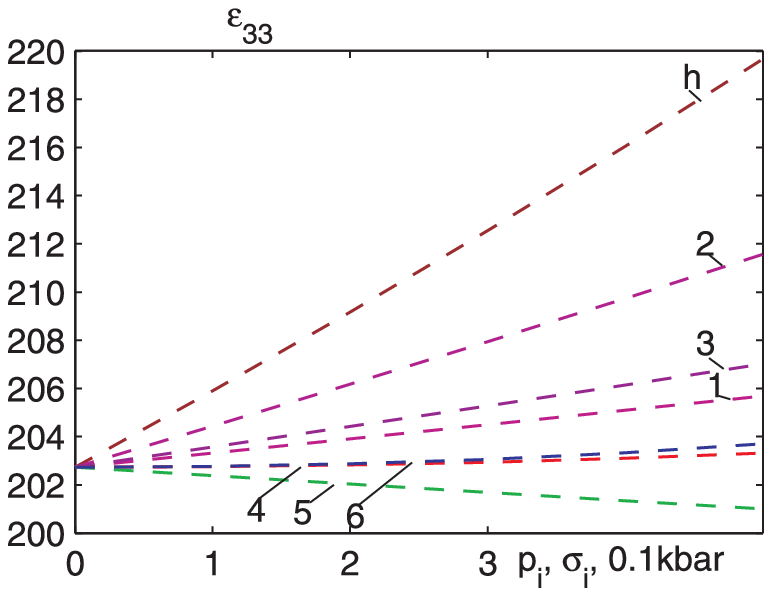}
\caption{(Colour online) Dependences of dielectric permittivity $\varepsilon_{33}$ of GPI crystal on hydrostatic pressure  (curve~h), uniaxial pressures ($p_{1}$~---~1, $p_{2}$~---~2, $p_{3}$~---~3) and shear stresses  ($\sigma_{4}$~---~4, $\sigma_{5}$~---~5, $\sigma_{6}$~---~6) at $\Delta T=-5$~K.} \label{eps33_p_sigma}
\end{minipage}
\end{figure}

\begin{figure}[!t]
\centering
\begin{minipage}{0.49\textwidth}
\centering
\includegraphics[width=0.95\textwidth]{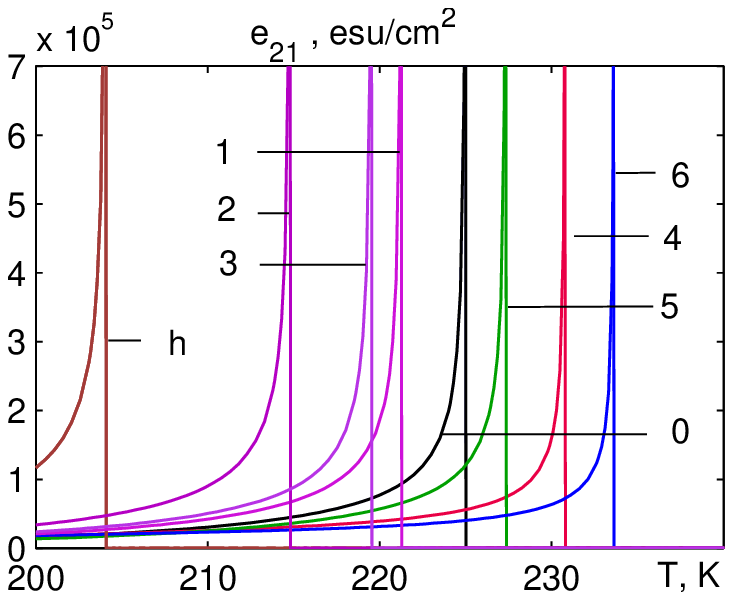}
\caption{(Colour online) The temperature dependences of coefficients of piezoelectric stress $e_{21}$ of GPI crystal   at zero stresses (curve~0),  under hydrostatic pressure (h), uniaxial pressures ($p_{1}$~---~1, $p_{2}$~---~2, $p_{3}$~---~3) and shear stresses   ($\sigma_{4}$~---~4, $\sigma_{5}$~---~5, $\sigma_{6}$~---~6). Value of the pressures and stresses is  2~kbar.} \label{e21_h123456}
\end{minipage}
\begin{minipage}{0.49\textwidth}
\centering
\vspace{-3mm}
\includegraphics[width=0.95\textwidth]{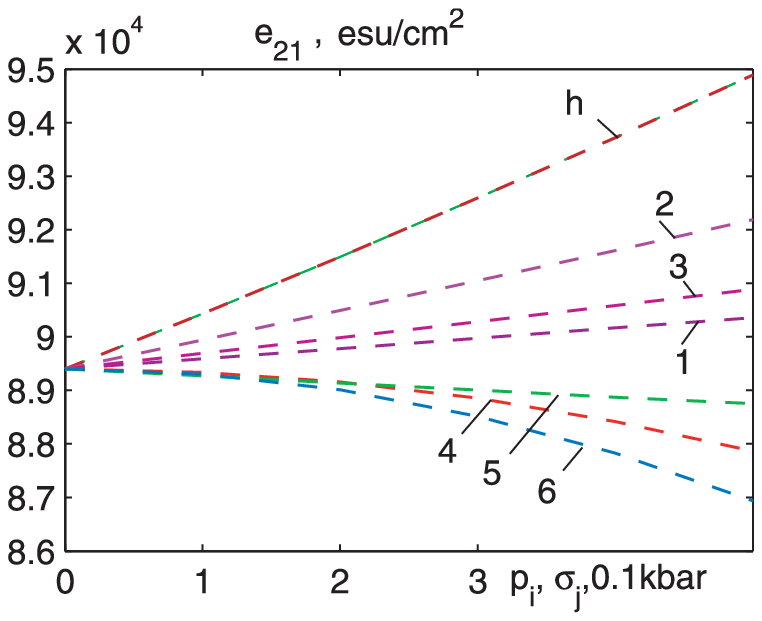}
\caption{(Colour online) Dependences of coefficients of piezoelectric stress $e_{21}$ of GPI crystal on hydrostatic pressure  (curve~h), uniaxial pressures ($p_{1}$~---~1, $p_{2}$~---~2, $p_{3}$~---~3) and shear stresses  ($\sigma_{4}$~---~4, $\sigma_{5}$~---~5, $\sigma_{6}$~---~6) at  $\Delta T=-4$~K.} \label{e21_p_sigma}
\end{minipage}
\end{figure}

\begin{figure}[!t]
\centering
\begin{minipage}{0.49\textwidth}
\centering
\includegraphics[width=0.95\textwidth]{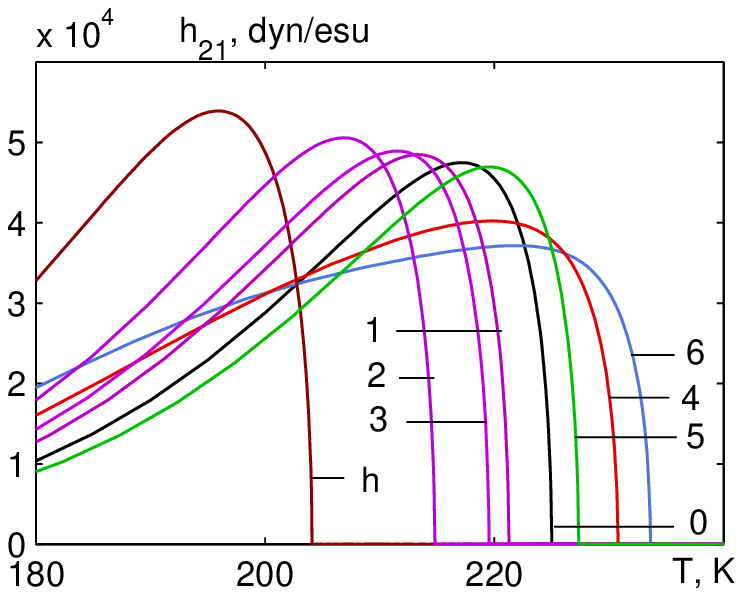}
\caption{(Colour online) The temperature dependences of constants of piezoelectric stress $h_{21}$ of GPI crystal   at zero stresses (curve~0),  under hydrostatic pressure (h), uniaxial pressures ($p_{1}$~---~1, $p_{2}$~---~2, $p_{3}$~---~3) and shear stresses   ($\sigma_{4}$~---~4, $\sigma_{5}$~---~5, $\sigma_{6}$~---~6). Value of the pressures and stresses is  2~kbar.} \label{h21_h123456}
\end{minipage}
\begin{minipage}{0.49\textwidth}
\centering
\vspace{-3mm}
\includegraphics[width=0.95\textwidth]{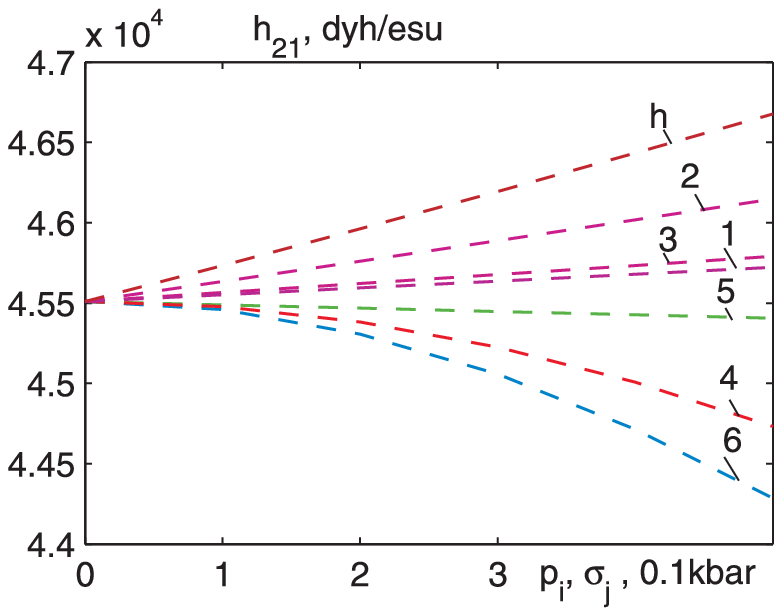}
\caption{(Colour online) Dependences of constants of piezoelectric stress $h_{21}$ of GPI crystal on hydrostatic pressure  (curve~h), uniaxial pressures ($p_{1}$~---~1, $p_{2}$~---~2, $p_{3}$~---~3) and shear stresses  ($\sigma_{4}$~---~4, $\sigma_{5}$~---~5, $\sigma_{6}$~---~6) at  $\Delta T=-4$~K.} \label{h21_p_sigma}
\end{minipage}
\end{figure}

\begin{figure}[!t]
\centering
\begin{minipage}{0.49\textwidth}
\centering
\includegraphics[width=0.95\textwidth]{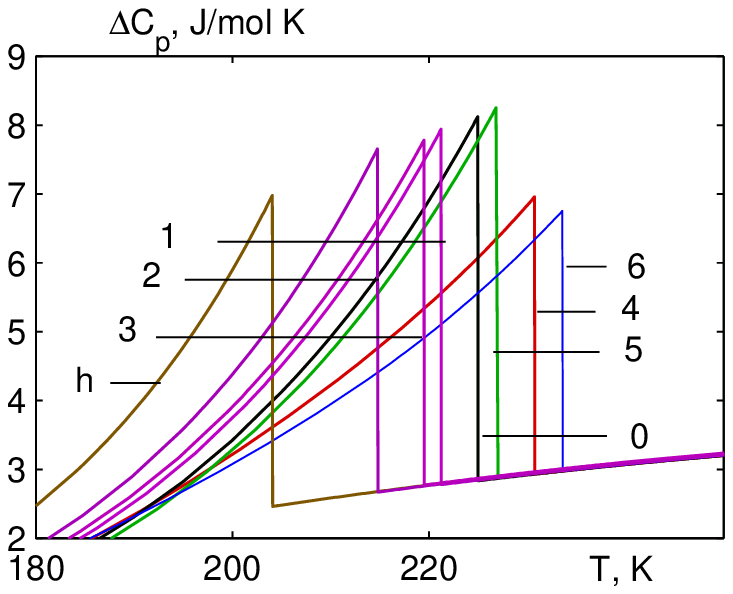}
\caption{(Colour online) The temperature dependences of proton contribution to molar heat capacity $\Delta C_{p}$ of GPI crystal   at zero stresses (curve~0),  under hydrostatic pressure (h), uniaxial pressures ($p_{1}$~---~1, $p_{2}$~---~2, $p_{3}$~---~3) and shear stresses   ($\sigma_{4}$~---~4, $\sigma_{5}$~---~5, $\sigma_{6}$~---~6). Value of the pressures and stresses is  2~kbar.} \label{Dcp_h123456}
\end{minipage}
\begin{minipage}{0.49\textwidth}
\centering
\vspace{-8mm}
\includegraphics[width=0.95\textwidth]{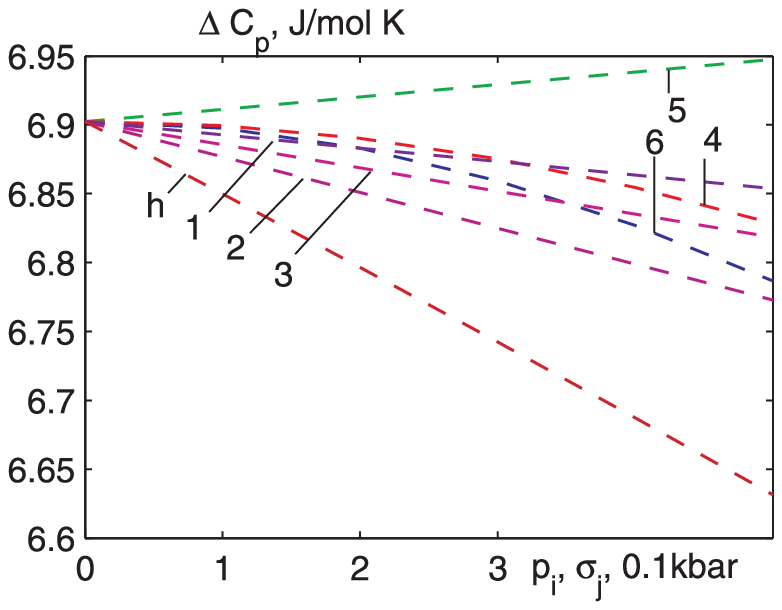}
\caption{(Colour online) Dependences of $\Delta C_{p}$ of GPI crystal on hydrostatic pressure  (curve~h), uniaxial pressures ($p_{1}$~---~1, $p_{2}$~---~2, $p_{3}$~---~3) and shear stresses  ($\sigma_{4}$~---~4, $\sigma_{5}$~---~5, $\sigma_{6}$~---~6) at $\Delta T=-5$~K.} \label{DCp_P_sigma}
\end{minipage}
\end{figure}

A decrease in the value of  $w_{1,2}$ and $\nu^+_{1, 2,3}$ under hydrostatic $p_{\text h}$ and uniaxial pressures  $p_{i}$ leads to a decrease of the phase transition temperature  $T_{\text c}$  of GPI crystal (figure~\ref{Tc_h123456}), 
and to a shift of the curves  h, 1, 2, 3 on the temperature dependences of spontaneous polarization $P_{2}$ (figure~\ref{P2_h123456}), components of dielectric permittivity  $\varepsilon_{22}$  (figure~\ref{eps22_h123456}), $\varepsilon_{11}$   (figure~\ref{eps11_h123456}), $\varepsilon_{33}$ (figure~\ref{eps33_h123456}), piezoelectric coefficients $e_{21}$ (figure~\ref{e21_h123456}),  $h_{21}$ (figure~\ref{h21_h123456}), molar heat capacity $\Delta C_{p}$  (figure~\ref{Dcp_h123456}).

With an increase of the pressures  $p_{\text h}$, $p_{i}$, the long-range interactions $\nu^+_{1, 2,3}$ decrease faster than the short-range interactions $w_{1}$ and $w_{2}$; that is, relation $w_{1,2}/\nu^+_{1, 2,3}$ increases. This leads to an increase with pressure of the above enumerated characteristics, excepting the heat capacity, at $\Delta T=\text{const}$ (curves~h, 1, 2, 3 and also h$'$, $1'$, $2'$, $3'$ on the pressure dependences in figures \ref{P2_p_sigma}, \ref{eps22_p_sigma}, \ref{eps11_p_sigma}, \ref{eps33_p_sigma}, \ref{e21_p_sigma}, \ref{h21_p_sigma}, \ref{DCp_P_sigma}).

The shear stress $\sigma_{5}$, on the contrary, leads to a slight increase of $w_{1,2}$, $\nu^+_{1, 2,3}$ and to a decrease of the relation  $w_{1,2}/\nu^+_{1, 2,3}$. As a result, the temperature $T_{\text c}$ increases, but the thermodynamic characteristics at $\Delta T=\text{const}$ slightly decrease  (curves 5, 5$'$ on the pressure dependences in figures \ref{P2_p_sigma}, \ref{eps22_p_sigma}, \ref{eps11_p_sigma}, \ref{eps33_p_sigma}, \ref{e21_p_sigma}, \ref{h21_p_sigma}).

The shear stresses   $\sigma_{4}$ and $\sigma_{6}$ influence the chain A (figure~\ref{struktura}) in the same way as $-\sigma_{4}$ and $-\sigma_{6}$ on the chain B. Therefore, the temperature $T_{\text c}$ and the thermodynamic characteristics do not depend on the sign of these stresses (curves 4, 6 on the pressure dependences are symmetric relative to 0).
The application of shear stresses  $\sigma_{4}$ and  $\sigma_{6}$ to the crystal leads to a crystal symmetry breakdown, and two sublattices  (chains A and  B) become nonequivalent.  Consequently, the parameters $w_{1}$ and $w_{2}$ split (curves $4_1$ and $4_2$ in figure~\ref{w12_p_sigma} instead of one curve 4, and also $6_1$ and $6_2$ instead of 6): interactions between pseudospins in the chain  A become stronger, while in  the chain  B they become weaker. Strengthening of interactions in a sublattice causes a phase transition in the crystal and increases temperature  $T_{\text c}$ (figure~\ref{Tc_h123456}, curves 4, 6), while the dependence $T_{\text c}(\sigma_{4, 6})$ is nearly hyperbolic cosine.
As a result, the temperature dependences  $P_{2}(T)$ (figure~\ref{P2_h123456}), $\varepsilon_{22}(T)$  (figure~\ref{eps22_h123456}), $e_{21}(T)$ (figure~\ref{e21_h123456}),  $h_{21}(T)$ (figure~\ref{h21_h123456}),  $\Delta C_{p}(T)$  (figure~\ref{Dcp_h123456}) shift to higher temperatures.

\begin{figure}[!t]
	\centering
		\includegraphics[scale=0.9]{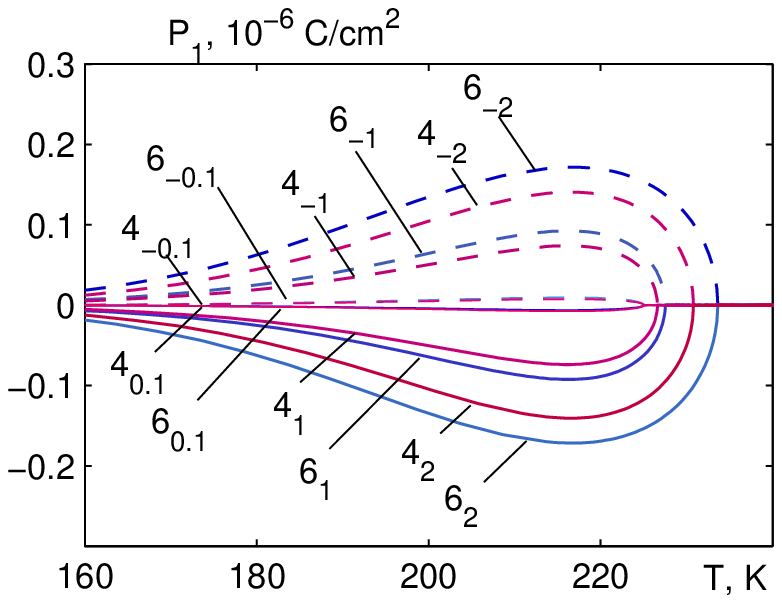}  \includegraphics[scale=0.9]{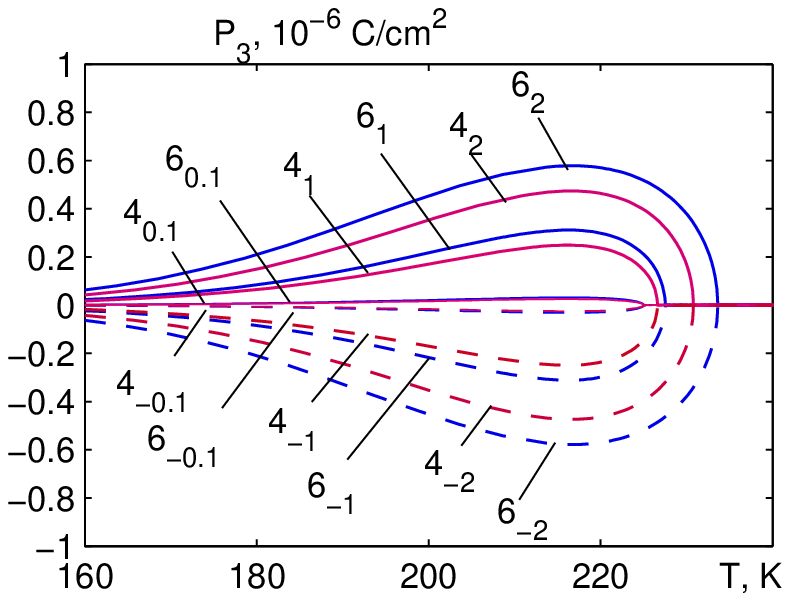}
	\caption[]{(Colour online) The temperature dependences of polarizations $P_{1}$ and $P_{3}$ of GPI crystal at different stresses~$\sigma_{4,6}$. Numbers of lines 4 and 6 mean the direction of the applied stress   $\sigma_{4}$ and  $\sigma_{6}$, respectively, the inferior index shows the value of the stresses~(kbar). }  \label{P1_p46_1}
\end{figure}

The influence of  shear stresses  $\sigma_{4}$ and  $\sigma_{6}$ in the plain  $XZ$ is qualitatively different. Since the chains~A and B become nonequivalent under these  stresses,  dipole moments of two sublattices  do not compensate each other in the plain  $XZ$ (analogously as in ferrimagnets). As a result, the components of spontaneous polarization $P_{1}$ and $P_{3}$ appear in the plane $XZ$ (figure~\ref{P1_p46_1}), and the curves $\varepsilon_{11}(T)$   (figure~\ref{eps11_h123456}, curves 4, 6), and $\varepsilon_{33}(T)$ (figure~\ref{eps33_h123456}) look like a longitudinal component of dielectric permittivity. 

\section{Conclusions}

In the present paper, the effects of hydrostatic pressure $p_{\text h}$,  uniaxial pressures $p_{1}$, $p_{2}$, $p_{3}$ and shear stresses  $\sigma_{4}$,  $\sigma_{5}$,  $\sigma_{6}$  on the phase transition and physical characteristics of this crystal are studied using the modified proton ordering model of GPI ferroelectric with hydrogen bonds by taking into account the piezoelectric coupling with strains
$\varepsilon_i$ in ferroelectric phase within two-particle cluster approximation.
Regularities of the change in strains   $\varepsilon_j$ under pressures $p_{\text h}$ and  $p_{i}$  and shear stresses $\sigma_{j}$ are determined. We have revealed that hydrostatic  $p_{\text h}$ and uniaxial  $p_{1}$ pressures weaken the short-range and long-range interactions in GPI crystal, while the long-range  interactions weaken more appreciably. As a result, the temperature $T_{\text c}$ almost linearly decreases, and  thermodynamic characteristics at $\Delta T=\text{const}$ increase.
In the case of $\sigma_{5}$, this is vice versa,  the temperature $T_{\text c}$ increases, and the values of thermodynamic characteristics at $\Delta T=\text{const}$ decrease.

Shear stresses  $\sigma_{4}$ and  $\sigma_{6}$, independently of the sign, lead to a nonlinear increase of the temperature~$T_{\text c}$. They influence the longitudinal characteristics similar to stress $\sigma_{5}$. However, due to a decrease of symmetry  and an incomplete compensation of dipole moments of two sublattices,  the transverse components of  polarization $P_{1}$ and $P_{3}$ appear in the plain $XZ$, and transverse permittivities $\varepsilon_{11}$ and $\varepsilon_{33}$ become similar to longitudinal permittivity $\varepsilon_{22}$.

For numerical calculations of thermodynamic characteristics under hydrostatic $p_{\text h}$ and uniaxial pressures~$p_{i}$ and under shear stresses $\sigma_{j}$,  we have not used additional model parameters, in comparison with the calculations  in the case of the absence of external influences. 
The temperature and pressure dependences of  thermodynamic characteristics of GPI crystal obtained in this work bear the character of predictions.

\section*{Acknowledgements}

This paper  is dedicated to a famous Ukrainian Professor Ihor Stasyuk on the occasion of his jubilee. Professor Stasyuk is a highly erudite scientist. He proposed methods and models that  have played a significant role in the construction of the microscopic theory of ferroelectrics during the recent five decades. The authors are sincerely thankful to  Professor Stasyuk for a valuable scientific discussion of the results of this work and useful advice and recommendations. We send our congratulations to Professor Stasyuk and wish him to remain in robust health, good luck and be happy.

\ukrainianpart

\title{Деформаційні ефекти в сегнетоелектрику фосфіт гліцину}
\author{ І.Р. Зачек\refaddr{label1}, Р.Р. Левицький\refaddr{label2}, А.С. Вдович\refaddr{label2}}
\addresses{\addr{label1} Національний університет ``Львівська політехніка'',  вул. С. Бандери, 12, 79013 Львів, Україна
	\addr{label2} Інститут фізики конденсованих систем НАН України,  вул. Свєнціцького, 1, 79011 Львів, Україна
}

\makeukrtitle

\begin{abstract}
\tolerance=3000%
	Для дослідження ефектів, що виникають під дією механічних напруг, використано модифіковану  модель сегнетоелектрика фосфіт гліцину
	шляхом врахування п'єзоелектричного зв'язку структурних елементів, які впорядковуються в цих кристалах, з
	деформаціями гратки. В наближенні двочастинкового кластера  розраховано компоненти векторa
	поляризації та тензора статичної діелектричної проникності механічно
	затиснутого  кристала, а також його п'єзоелектричні  та теплові характеристики.  Досліджено вплив зсувних напруг, гідростатичного та одновісних тисків  на фазовий перехід та фізичні характеристики кристалу.

	\keywords сегнетоелектрики, фазовий перехід, діелектрична проникність, п'єзоелектричні коефіцієнти, механічна напруга
\end{abstract}

\end{document}